\documentclass[english,superscriptaddress]{revtex4-2}
\usepackage[T1]{fontenc}
\usepackage[latin9]{inputenc}
\usepackage{amsmath}
\usepackage{amssymb}
\usepackage{graphicx}
\usepackage{color}

\makeatletter

\providecommand{\tabularnewline}{\\}

\makeatother

\usepackage{babel}
\begin{document}
\title{Bubble nucleation and gravitational wave from holography}
\author{Yidian Chen}
\email{chenyidian@ucas.ac.cn}

\affiliation{School of Nuclear Science and Technology, University of Chinese Academy
of Sciences, Beijing 100049, P.R. China}
\author{Danning Li}
\email{lidanning@jnu.edu.cn}

\affiliation{Department of Physics and Siyuan Laboratory, Jinan University, Guangzhou
510632, P.R. China}
\author{Mei Huang}
\email{huangmei@ucas.ac.cn}

\affiliation{School of Nuclear Science and Technology, University of Chinese Academy
of Sciences, Beijing 100049, P.R. China}
\begin{abstract}
We investigate the bounce solution in the holographic QCD and electroweak models with first-order phase transition. 
The strength parameter $\alpha$, inverse duration time $\beta/H$ and bubble wall velocity $v_w$ in the gravitational 
wave power spectra are calculated by holographic bounce solution. In contrast to the results of field theory, 
we find the parameter $\alpha$ is about $\mathcal{O}(1)$ and $\beta/H$ is about $10^4$, which imply that the phase transition is fast and strong. 
The critical, nucleation and percolation temperatures of the phase transition are close to each other in the holographic model. 
In addition, the velocity $v_w$ is found to be less than the sound speed of the plasma $c_{s}=1/\sqrt{3}$, which corresponds to the deflagration scenario.
For QCD phase transition, the gravitational wave power spectrum can reach $10^{-13}-10^{-14}$ around the peak frequency of 0.01 Hz, 
which can be detected by BBO and Ultimate-DECIGO. For electroweak phase transition, the gravitational wave power spectrum can reach $10^{-12}-10^{-16}$ around the peak frequency $1-10$ Hz. 
Moreover, the primordial black hole is not favorable for formation due to the large parameter $\beta/H$ and small velocity $v_w$.
\end{abstract}
\maketitle

\section{introduction}

Gravitational waves (GWs) are one of the great predictions of general
relativity \citep{einstein1916,einstein1918}, which exhibits an effect
of the curvature of spacetime. In 1974, Hulse and Taylor discovered
the binary system PSR 1913+16 \citep{Hulse:1974eb}, providing indirect
evidence for the existence of GWs, for which they received the 1993
Nobel Prize in Physics. In 2015, astronomy entered the multi-messenger
era with the first observation of a gravitational wave event by LIGO
\citep{LIGOScientific:2016aoc,LIGOScientific:2021djp}. Nowadays,
more and more GW events or possible GW signals are discovered, as
in Refs. \citep{NANOGrav:2020bcs}, which also provide new instruments
for understanding cosmology and astronomy deeply. 

The GWs come from the quadrupole moment radiation of the stress-energy
tensor, which can be roughly classified into two categories, that
is, cosmological and astronomical sources (see the reviews Refs. \citep{Maggiore:2007ulw,Cai:2017cbj,Maggiore:2018sht}).
The stochastic GW background generated by the first-order
phase transition (FOPT) of the early universe is a significant cosmological
source of GWs (see the reviews Ref. \citep{Christensen:2018iqi}).
Different from the transient GWs observed at the present,
the stochastic GW background comes from all directions
rather than specific ones.

The detection experiments of GWs mainly include ground-based experiments,
space-based experiments, pulsar timing arrays (PTA), and cosmic microwave
background polarization, (see \citep{Bailes:2021tot} for review).
The ground-based experiments, such as LIGO \cite{LIGOScientific:2014pky}, Virgo \citep{VIRGO:2014yos}, Einstein
Telescope (ET) \citep{Punturo:2010zz}, Cosmic Explorer (CE) \citep{LIGOScientific:2016wof}, etc., mainly observe compact binary systems. The space-based
experiments, such as Laser Interferometer Space Antenna (LISA) \cite{LISA:2017pwj}, Deci-Hertz
Interferometer Gravitational wave Observatory (DECIGO) \citep{Kawamura:2006up,Kudoh:2005as}, Big Bang Observer
(BBO) \citep{Harry:2006fi}, Taiji \citep{Hu:2017mde}, Tianqin \citep{TianQin:2015yph}, etc., are more sensitive to the GWs from electroweak
phase transition. The PTAs, such as the Parkes PTA (PPTA) \citep{Manchester:2012za}, the European PTA
(EPTA) \citep{Kramer:2013kea}, the North American Nanohertz Observatory for Gravitational Waves (NANOGrav) \citep{McLaughlin:2013ira}, the International PTA (IPTA) \citep{Manchester:2013ndt} and the Chinese PTA (CPTA), etc., are mainly sensitive to the GWs from QCD
phase transitions.

The dynamic process of the FOPT is described
by bubble dynamics. When the temperature of the system reaches the
critical temperature $T_{c}$ for the phase transition, it does not
immediately enter the symmetric broken phase since the generation of
true vacuum bubbles causes additional surface free energy, which increases
the total free energy of the system. As the temperature decreases
to the nucleation temperature $T_{n}$, bubbles are more likely to
be generated due to the increasing probability of thermal perturbation
or quantum tunneling that crosses the free energy barrier. The temperature
drops further to the percolation temperature $T_p$, and more than one-third
of the space enters into the true vacuum. During this process, the
bubble expands continuously since the internal pressure is greater
than the false vacuum pressure and surface tension, and it may eventually
reach the final velocity because of the friction of the plasma or
accelerate to near the speed of light. Throughout the process, collisions
of bubbles, acoustic modes of the plasma and turbulences of the
magnetohydrodynamics all generate GWs and eventually contribute to
the GW power spectra.

During the evolution of the universe, it may undergo various phase
transitions, such as grand unification phase transition, electroweak phase transition (EWPT), and QCD
phase transition (QCDPT). The FOPT is of
interest since it is related to some physical processes such as baryogenesis,
the seeds of intergalactic magnetic fields and the formation of
primordial black holes. Unfortunately, the electroweak part of the
standard model is crossover \citep{Kajantie:1996mn,Gurtler:1997hr,Csikor:1998eu},
while the Lattice QCD calculations indicate that the QCD phase transition
of the three flavors is crossover at zero chemical potential and finite
temperature \citep{Fodor:2001au,Ding:2015ona}. Of course, many new physical models beyond the standard
model (BSM) predict the FOPT, such as the two-Higgs
doublet model \citep{Cline:1996mga,Basler:2016obg,Dorsch:2016nrg}, the left-right symmetric model \citep{Li:2020eun}, the technicolor model \citep{Appelquist:1996dq,Sannino:1999qe,Appelquist:1999dq},
etc. In addition, since QCDPT is flavor-dependent, there is still
the possibility of a first-order QCDPT.

The discovery of the anti-de Sitter/conformal field theory (AdS/CFT)
correspondence \citep{Maldacena:1997re,Gubser:1998bc,Witten:1998qj}
has provided a new way to solve strongly coupled field theory calculations.
In the past two decades, holographic QCD has been widely studied both in
top-down \citep{Erdmenger:2007cm,Sakai:2004cn,Sakai:2005yt} and bottom-up models \citep{Erlich:2005qh,Karch:2006pv,Gursoy:2007cb,Gubser:2008ny,Grefa:2021qvt,Li:2013oda,Chen:2022goa}. For beyond the standard model the technicolor
model \citep{Haba:2008nz,Matsuzaki:2012xx,Elander:2012fk,Chen:2017cyc,BitaghsirFadafan:2018efw,Chen:2019aku} and composite Higgs model \citep{Contino:2003ve,Agashe:2004rs,Croon:2015wba,Espriu:2017mlq} have been extended to the holographic framework.

The intensity of the GW signal is impacted by the strength of the phase transition, its duration time and the final velocity of the bubble wall. 
The calculations in weakly coupled quantum field theory suggest that the strength parameter $\alpha$ is roughly $\alpha\sim 0.01$ and the inverse of the duration time $\beta/H$ is roughly $\beta/H\sim 100$ during EWPT. As for the speed, one would expect that it is close to the speed of light $c$ for enhancing the GW signal. These quantities are not sufficiently discussed and understood when the system is strongly coupled. The holographic principle provides new ways to explore the thermodynamic and kinetic properties of phase transitions with strong coupling. 
The bubble nucleation dynamics \citep{Attems2018a,Attems2020,Bea2021a,Bea:2021zol,Bigazzi2021,Bigazzi2021a,Ares2022,Ares2022a,Bea2022,Bea2022a,Chen:2022cwi,Janik2022a} and the GW power spectra \citep{Ahmadvand:2017xrw,Ahmadvand:2017tue,Chen:2017cyc,Rezapour:2020mvi,Zhu:2021vkj,Novikov:2022lqf,Cai:2022omk,He:2022amv} are considered in the holographic model.
The relation between
the bubble velocity and the pressure difference of the true and false
vacuum is investigated in Refs. \citep{Bea2021a,Bigazzi2021,Janik2022a}.
In Refs. \citep{Bea2022a,Bea2022}, the profile of the fluid velocity
is calculated inside and outside the bubble wall.

The paper is organized as follows. The five-dimensional holographic QCD and electroweak models are introduced in Sec. \ref{sec:2}. In Sec. \ref{sec:3}, the bounce solution and bubble wall velocity are obtained in the holographic model. Using the bounce solution, the properties of the critical bubble and the thin-wall approximation are considered in Sec. \ref{sec:4}. In Sec. \ref{sec:5}, the stochastic GW power spectra generated by the strongly coupled FOPT are calculated. Finally, the conclusion and discussion are presented in Sec. \ref{sec:6}.

\section{5d setup}
\label{sec:2}

In this section, we consider FOPT of
QCD or QCD-like electroweak theories, which corresponds to $SU(N_{f})_{L}\times SU(N_{f})_{R}$
flavor symmetry breaking to $SU(N_{f})_{V}$ subgroup. Here, only
the scalar part of the flavor-brane is considered under the probe
approximation, which has the following form
\begin{equation}
S=-\int d^{5}x\sqrt{-g}e^{-\Phi}{\rm Tr}[(D^{M}X)^{\dagger}(D_{M}X)+V_{X}(|X|)].\label{eq:action}
\end{equation}
Among this, $\Phi$ denotes the dilaton field and the complex scalar
field $X$ corresponds to the quark condensation $\langle\bar{q}q\rangle$
or fermionic condensate of BSMs. The symmetry of the model is spontaneously
broken by the non-zero vacuum expectation value of the scalar field
$X=\frac{\chi(z,x)}{\sqrt{2N_{f}}}I_{N_{f}}$, where $I_{N_{f}}$
is the $N_{f}\times N_{f}$ identity matrix and $\chi$ depends not
only on the fifth coordinate $z$ but also on the four-dimensional
space-time coordinates $x^{\mu}$.

In this paper, the back-reaction of the dilaton field $\Phi$ and
the scalar field $X$ are not taken into account, so the background
geometry remains the ${\rm AdS_{5}}$-Schwarzchild black brane metric
\begin{equation}
ds^{2}=\frac{L^{2}}{z^{2}}[-f(z)dt^{2}+\frac{1}{f(z)}dz^{2}+dx_{i}dx^{i}],
\end{equation}
with the blackening factor $f(z)=1-\frac{z^{4}}{z_{h}^{4}}$ and the
horizon $z_{h}$. For convenience, the AdS radius $L$ is set to 1
in the following. The Hawking temperature of the system is
\begin{equation}
T=\frac{|f^{'}(z)|}{4\pi}=\frac{1}{\pi z_{h}}.
\end{equation}

In order to realize the FOPT, the potential
of the scalar field is considered to have the following form\citep{Chelabi:2015cwn,Chelabi:2015gpc,Chen:2019rez}
\begin{equation}
V(\chi)\equiv{\rm Tr}[V_{X}(|X|)]=\frac{M_{5}^{2}}{2}\chi^{2}+\upsilon_{3}\chi^{3}+\upsilon_{4}\chi^{4}+\upsilon_{6}\chi^{6},
\end{equation}
where $M_{5}^{2}$ denotes the square of the five-dimensional mass,
and $\upsilon_{3}$, $\upsilon_{4}$ and $\upsilon_{6}$ are the cubic,
quadratic and sextic terms coupling constants, respectively. According
to the AdS/CFT dictionary, the five-dimensional mass of the scalar
field $X$ is $M_{5}^{2}=(\Delta-p)(\Delta+p-4)=-3$ by taking $p=0$
and $\Delta=3$ for QCD. For the EW model, a large anomalous dimension
$\gamma_{m}\simeq1$ needs to be considered\citep{Appelquist:1991is,Sundrum:1991rf,Appelquist:1998xf,Harada:2005ru,Kurachi:2006mu,Kurachi:2007at},
so the five-dimensional mass is $M_{5}^{2}=(\Delta-\gamma_{m})(\Delta-\gamma_{m}-4)=-4$
by taking $\Delta=3$ and $\gamma_{m}=1$. For simplicity, we do not
consider all nonzero nonlinear terms, but instead consider two special
cases, i.e., $\upsilon_{6}=0$ (Model I) or $\upsilon_{3}=0$ (Model II). The correct chiral
symmetry breaking in the chiral limit depends on the form of the dilaton
field, which is chosen as in Refs. \citep{Chelabi:2015cwn,Chelabi:2015gpc}
\begin{equation}
\Phi=-\mu_{1}z^{2}+(\mu_{1}+\mu_{0})z^{2}\tanh(\mu_{2}z^{2}),
\end{equation}
where the $\mu_{0}$, $\mu_{1}$, and $\mu_{2}$ parameters determine
the behavior of the dilaton field in the IR and UV. 

\begin{figure}
\includegraphics[width=0.4\textwidth]{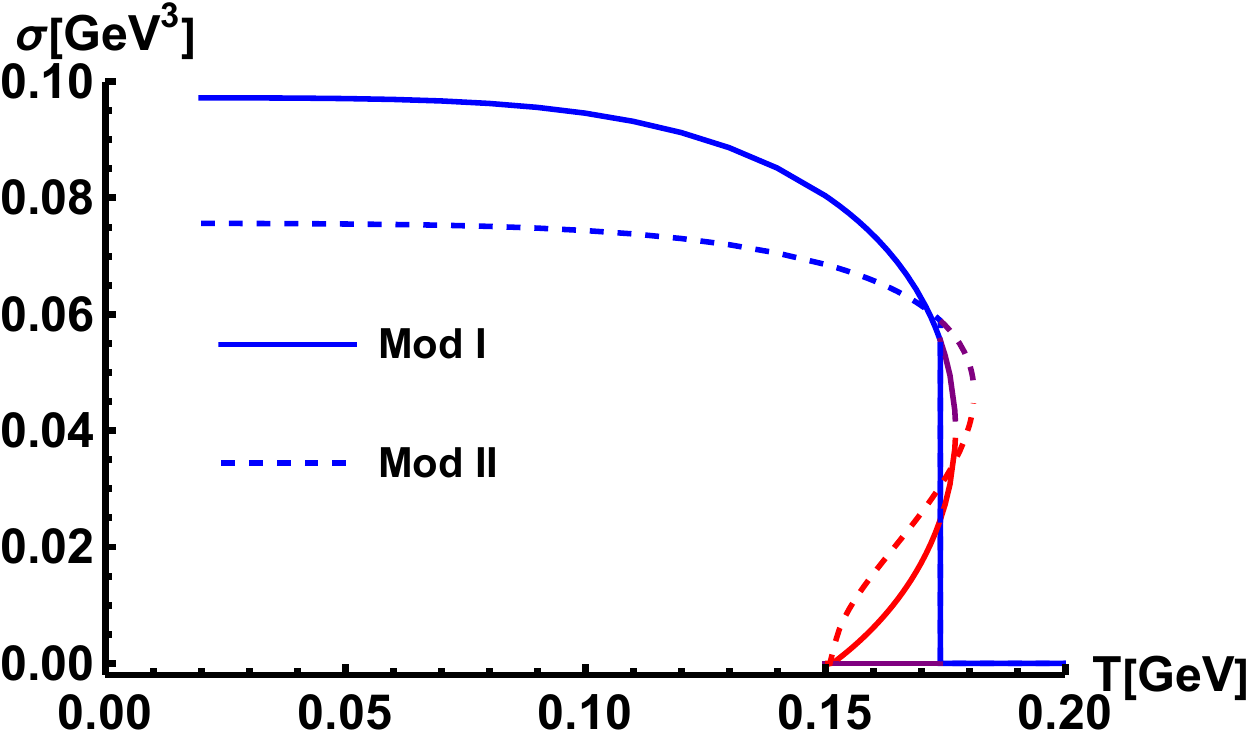}\hspace{1.5cm}\includegraphics[width=0.4\textwidth]{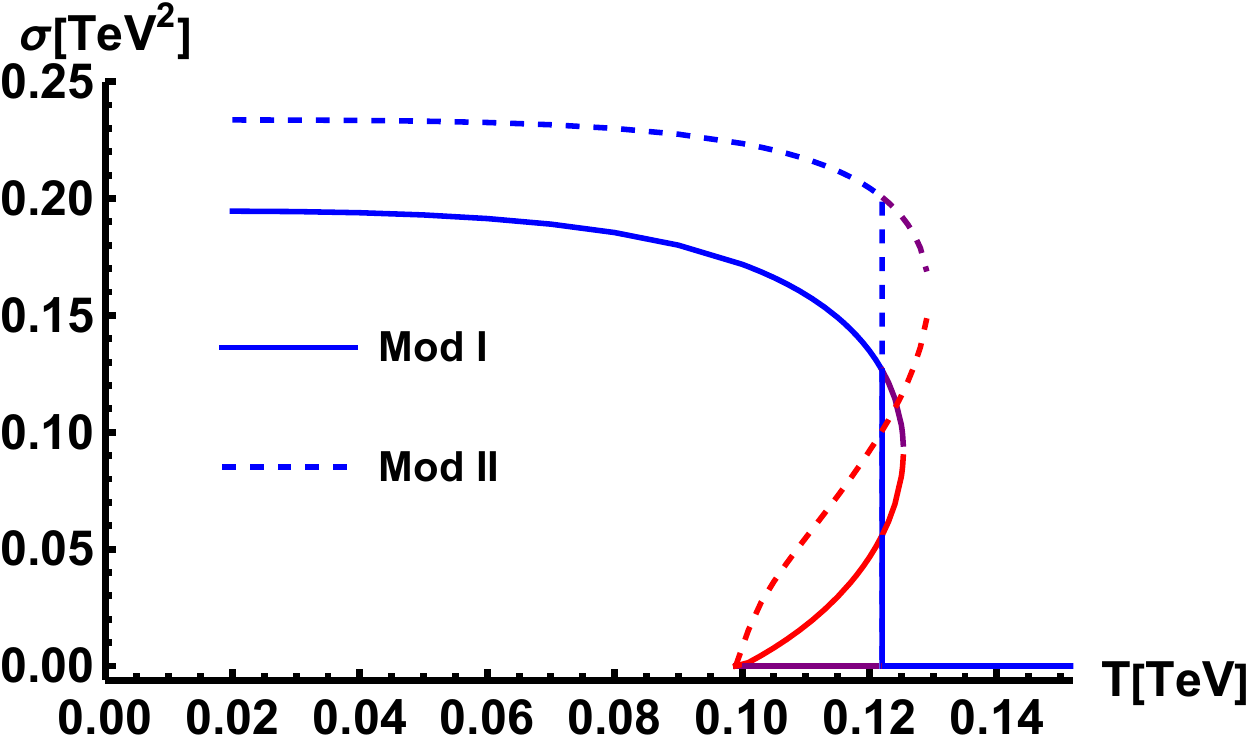}
\vskip -0.05cm \hskip 0 cm
\textbf{( a ) } \hskip 8 cm \textbf{( b )}
\caption{\label{fig:sigma_T}The condensate $\sigma$ as a function of temperature
$T$. Panels (a) and (b) correspond to the QCD and EW cases, respectively.}
\end{figure}

In this holographic model, five free parameters $[\mu_{0},\mu_{1},\mu_{2},\upsilon_{3},\upsilon_{4}]$
or $[\mu_{0},\mu_{1},\mu_{2},\upsilon_{4},\upsilon_{6}]$ are included.
For the QCD case, we choose $[\mu_{0},\mu_{1},\mu_{2},\upsilon_{3},\upsilon_{4}]=[(0.43{\rm GeV})^{2},(0.83{\rm GeV})^{2},(0.176{\rm GeV})^{2},-3,8]$ (Model I) 
or $[\mu_{0},\mu_{1},\mu_{2},\upsilon_{4},\upsilon_{6}]=[(0.43{\rm GeV})^{2},(0.83{\rm GeV})^{2},(0.176{\rm GeV})^{2},-12,100]$ (Model II) 
with the critical temperature of the phase transition $T_{c}\simeq174$
MeV, by referring to Refs. \citep{Chelabi:2015cwn,Chelabi:2015gpc,Chen:2019rez}.
For the EW case, $\mu_{0}=\mu_{2}=0$ is chosen to simplify the model.
It is worth noting that if $\Phi=-\mu_{1}z^{2}$ is chosen, there
is an additional massless scalar meson in the particle spectra, as in
Ref. \citep{Karch2011}. For QCD, this is nonphysical. However, for
the EW, this state can be interpreted as the Higgs boson, which obtains
mass by interacting with other scalar fields. Therefore,
for the EW case, the parameters are chosen as $[\mu_{0},\mu_{1},\mu_{2},\upsilon_{3},\upsilon_{4}]=[0,(0.28{\rm TeV})^{2},0,-1.1,3]$ (Model I)
or $[\mu_{0},\mu_{1},\mu_{2},\upsilon_{4},\upsilon_{6}]=[0,(0.28{\rm TeV})^{2},0,-1.8,4]$ (Model II) 
with critical temperatures of $T_{c}\simeq122$ GeV. The Panels. (a)
and (b) of Fig. \ref{fig:sigma_T} show the QCD and technicolor condensate
as a function of temperature, respectively. In fact, the specific
values of these parameters do not affect the qualitative results in
the following sections. In principle, the same approach can be used
for symmetry breaking of other groups, simply by changing the representation
of the group.

\section{bounce solution and bubble velocity}
\label{sec:3}

In this section, we will study the kinetics of the FOPT in the holographic model, which will involve the thermodynamics
and dynamics of the bubble, describing how the phase transition occurs
and how long the phase transition lasts. The phase transition kinetics
are characterized by nucleation temperature, latent heat, transition
rate parameters, bubble velocity, etc., which will determine the specific
magnitude of the GW spectrum.

The kinetic process of the FOPT is described
by the nucleation theory, in which the creation, expansion and fusion
of bubbles transform the false vacuum $\chi_{f}$ into the true vacuum
$\chi_{t}$. The bubble nucleation is caused by quantum tunneling
or thermal perturbation. At the critical temperature $T_{c}$, the
true and false vacuums have the same free energy, and the phase transition
is suppressed due to the surface free energy of the bubbles increasing
the total free energy of the system. As the temperature decreases,
the difference between the free energy of the true and false vacuum
compensates for the surface free energy, and the bubble generation
becomes more and more probable until the phase transition is completed.

In the first half of the 20th century, classical and modern nucleation
theories were established\citep{becker1935kinetische,Langer:1969bc}.
In the 1970s, nucleation theory was extended to relativistic quantum
field theory by Coleman and Callan\citep{Coleman:1977py,Callan:1977pt},
while in Refs. \citep{Affleck:1980ac,Linde:1980tt,Linde:1981zj,Csernai:1992tj}
and \citep{Venugopalan:1993vk} it was further extended to finite
temperature and density, respectively. See Refs. \citep{mcdonald1962homogeneous,mcdonald1963homogeneous,kalikmanov2013nucleation}
for some reviews.

\subsection{Bounce Solution}

In this subsection, we will construct bubble solutions in the holographic
model. Considering the spherical symmetry of the bubble, the AdS-Schwarzchild
metric is rewritten in spherical coordinates
\begin{equation}
ds^{2}=\frac{1}{z^{2}}[-f(z)dt^{2}+\frac{1}{f(z)}dz^{2}+dr^{2}+r^{2}d\theta^{2}+r^{2}\sin^{2}\theta d\varphi^{2}],\quad(0\leq r\leq R)
\end{equation}
with azimuth $\varphi$, zenith angle $\theta$ and the edge $R$. Then, the equations
of motion of the scalar field $X$ can be obtained from action (\ref{eq:action})
\begin{equation}
\partial_{r}^{2}\chi(z,r)+\frac{2\partial_{r}\chi(z,r)}{r}+f(z)\partial_{z}^{2}\chi(z,r)+\left(f'(z)-f(z)\Phi'(z)-\frac{3f(z)}{z}\right)\partial_{z}\chi(z,r)-\frac{\partial_{\chi}V(\chi)}{z^{2}}=0.\label{eq:eom}
\end{equation}

Coleman and Callan\citep{Coleman:1977py,Callan:1977pt} proposed that
the critical bubble is described by the bounce solution in quantum
nucleation theory. The solution requires the following boundary conditions
\begin{eqnarray}
\lim_{\text{\ensuremath{r\to\infty}}}\chi(z,r) & = & \chi_{f},\label{eq:bc-1}\\
\frac{d\chi(z,r)}{dr}\Bigg|_{r=0} & = & 0.\label{eq:bc-2}
\end{eqnarray}
For the fifth dimensional direction, the expansion of the scalar field
$\chi$ at the conformal boundary has the following form
\begin{eqnarray}
\chi|_{z\to0} & = & m_{q}\zeta z+...+\frac{\sigma}{\zeta}z^{3}+...\quad(\text{QCD}),\label{eq:bc-3}\\
 & = & \sigma z^{2}+...\quad(\text{EW}),\label{eq:bc-4}
\end{eqnarray}
with the quark current mass $m_{q}$, the condensate $\sigma$ and
constant $\zeta=\frac{\sqrt{N_c}}{2\pi}$. Here, we keep the terms related with the 4D operators only, and in the following calculation we will take $N_c=3$ for the QCD case.  Since the FOPT is considered, it is convenient to set the current
mass $m_{q}$ to 0. At the IR boundary, the natural boundary condition
is selected. Under saddle point approximation, the tunneling rate
of the stochastically generated bubbles is
\begin{equation}
\Gamma(T)=A(T)e^{-\frac{S_{b}}{T}},\label{eq:bubble_p}
\end{equation}
where $S_{b}$ is the Euclidean action evaluated on the bounce solution
and the factor is $A=T^{4}(\frac{S_{b}}{2\pi T})^{3/2}$\citep{Linde:1980tt,Linde:1981zj}. 
According to the holographic principle, the partition function has the equivalence $\mathcal{Z}_{\rm QFT}\simeq \mathcal{Z}_{\rm Gra}$, so the action $S_b$ can be calculated by the gravitational part $S_b\simeq S_5$.
The Euclidean action of the bounce solution has the following form

\begin{eqnarray}
S_{b} & \simeq S_5 = & 4\pi\int_{0}^{R}dr\int_{0}^{z_{h}}dz\sqrt{-g}e^{-\Phi(z)}(-\frac{\upsilon_{3}}{2}\chi^{3}-\upsilon_{4}\chi^{4}-2\upsilon_{6}\chi^{6}).\label{eq:onshellaction}
\end{eqnarray}

\begin{figure}
\includegraphics[width=0.29\textwidth]{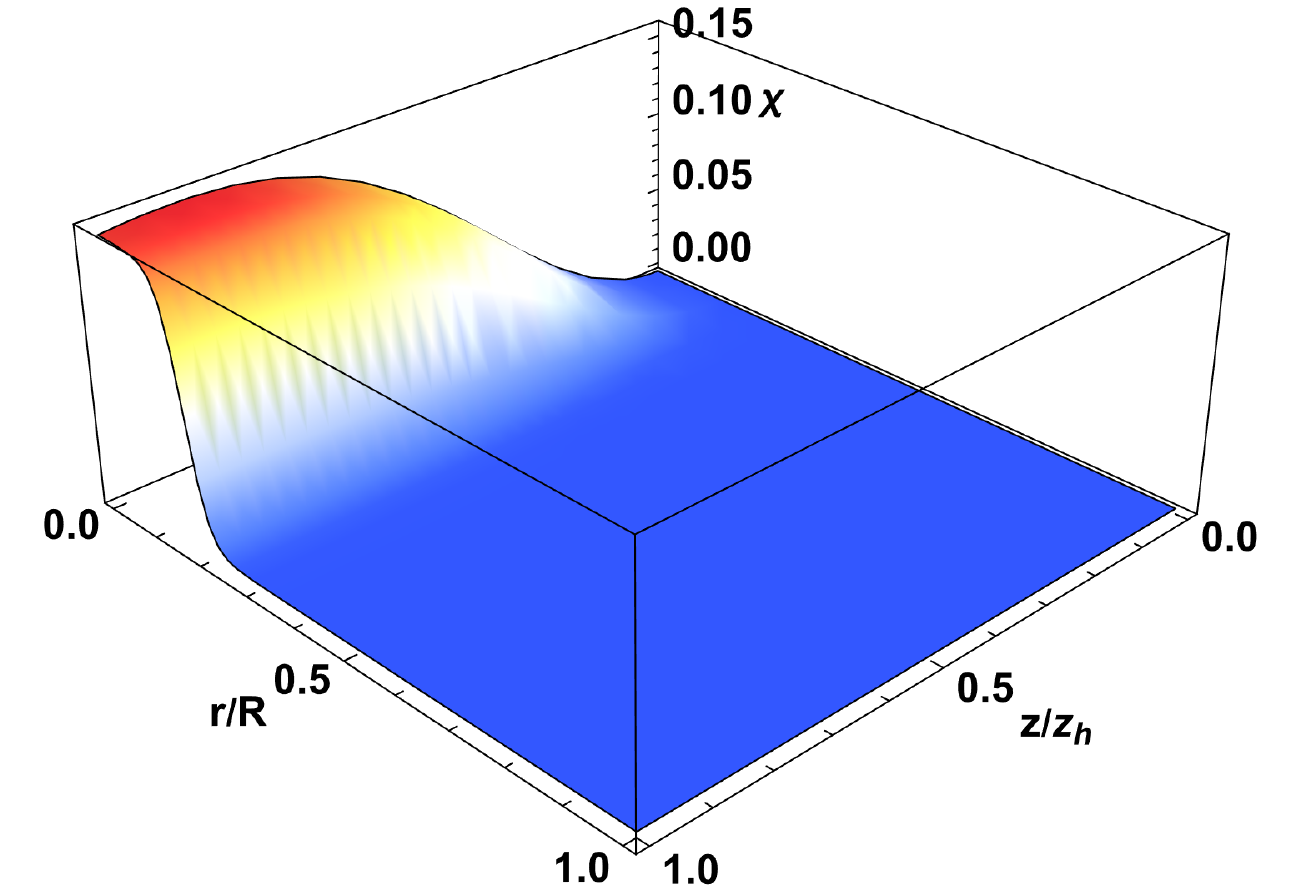}\hspace{1.5cm}\includegraphics[width=0.48\textwidth]{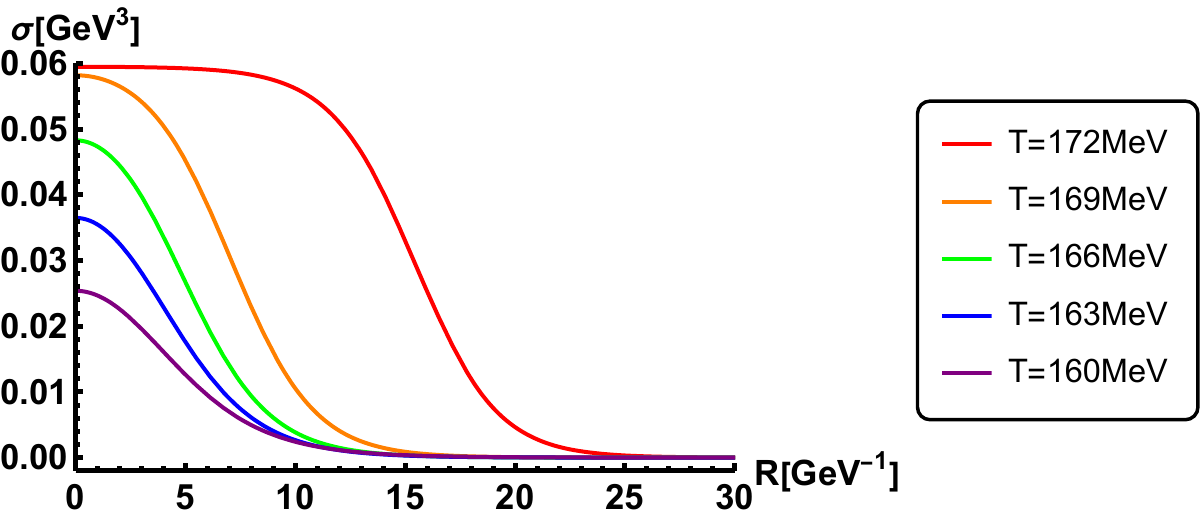}
\vskip -0.05cm \hskip 0 cm
\hskip -3 cm \textbf{( a ) } \hskip 6 cm \textbf{( b )}
\caption{\label{fig:bubble_T}The panel (a) represents the bounce solution
of the scalar field $\chi_b$ as a function of the fifth dimensional
coordinate $z$ and the radial coordinate $r$. The panel (b) indicates
the condensate $\sigma$ as a function of the radial coordinate
$r$ at different temperatures.}
\end{figure}

By solving the equation of motion Eq. (\ref{eq:eom}) with the boundary
conditions Eqs. (\ref{eq:bc-1}-\ref{eq:bc-4}),
we can obtain the bounce solution in the holographic model. For simplicity,
we focus on the case of the QCD phase transition with $\upsilon_{3}\neq0$ (Model I).
For other parameter values or holographic EW models, the qualitative conclusions
do not change. The panel (a) of Fig. (\ref{fig:bubble_T}) shows the
scalar field $\chi$ as a function of the fifth dimensional coordinate
$z$ and the radial coordinate $r$ at a temperature of 172 MeV. It
can be seen that the scalar field $\chi$ has a nontrivial structure
as a function of $z$ when the radial $r$ is small, while the profile
of $\chi$ varies continuously to the trivial solution as $r$ tends
to the edge $R$. 

The panel (b) of Fig. \ref{fig:bubble_T} represents the condensation
as a function of radial $r$ at different temperatures. It can be
seen that the size of the critical bubble diminishes with decreasing
temperature. As the temperature decreases, the free energy barrier between the true and false vacuums is reduced, and small-size bubbles are more likely to form. 
It should be noted that
the condensation values at the center of the bubble do not reach the equilibrium values
when $T\lesssim170$ MeV. This can be interpreted as the fact that at that temperature, the bubbles are composed mainly of bubble walls, as can be seen from panel (b) of Fig. \ref{fig:bubble_T}. 
The true vacuum is revealed inside the bubble when $T\gtrsim170$ MeV.

\subsection{Bubble velocity}

When a large enough bubble is created in the plasma, it expands, collides
with each other and eventually transforms the false vacuum into a
true vacuum. During this process, some physical phenomena such as
baryogenesis, GW generation, and primordial black hole formation will
happen. It can be expected that the velocity of the bubble wall has
an impact on the final signal. For the GW power spectrum, numerical
simulations indicate that faster bubble wall velocity enhances the
signal intensity \citep{Caprini:2015zlo}.

The velocity of the bubble wall in the plasma is governed by hydrodynamics
and particle interactions. The way bubbles expand is divided into
deflagration, detonation, hybrid and runaway cases \citep{Espinosa:2010hh},
which influence the GW spectra through bubble collisions and interactions
with the hydrodynamics. In order to understand the details of FOPT
more accurately, many methods have been used to calculate the bubble
wall velocity. As in Refs. \citep{Konstandin:2010dm,BarrosoMancha:2020fay,Balaji:2020yrx,Ai:2021kak},
the local equilibrium hydrodynamic equations are applied to obtain
the velocity. More generally, considering the effect of out of equilibrium,
people need to solve the distribution function by the Boltzmann equation,
which allows to obtain the velocity containing the backreaction \citep{Moore:1995si,Moore:1995ua,Konstandin:2014zta,Kozaczuk:2015owa,Laurent:2020gpg,Dorsch:2021nje,DeCurtis:2022hlx,Laurent:2022jrs}.
In addition, the holographic method is also applied to the calculation
of bubble wall velocity. In Refs. \citep{Bea2021a,Bea:2021zol,Bigazzi2021,Bea2022a,Bea2022,Janik2022a}
, applying the gauge/gravity duality, the bubble velocity is calculated
for strongly coupled hydrodynamics.

In this letter, the probe approximation is considered, i.e., we only
account for the obstruction from the plasma fluid without the backaction.
To investigate the real-time evolution, we transform the framework
to the ingoing Eddington-Finkelstein coordinate, then the metric becomes
\begin{eqnarray}
ds^{2} & = & \frac{1}{z^{2}}[-f(z)dt^{2}-2dtdz+dr^{2}+r^{2}d\theta^{2}+r^{2}\sin^{2}\theta d\varphi^{2}].
\end{eqnarray}
Under the coordinate transformation, the scalar field $\chi$ is invariant,
and its equation of motion becomes as follows
\begin{eqnarray}
\partial_{r}^{2}\chi(t,z,r)+\frac{2\partial_{r}\chi(t,z,r)}{r}+f(z)\partial_{z}^{2}\chi(t,z,r)+\left(f'(z)-f(z)\Phi'(z)-\frac{3f(z)}{z}\right)\partial_{z}\chi(t,z,r)\nonumber \\
-\frac{\partial_{\chi}V(\chi)}{z^{2}}+\left(\frac{3}{z}+\Phi'(z)\right)\partial_{t}\chi(t,z,r)-2\partial_{t}\partial_{z}\chi(t,z,r) & = & 0.\label{eq:eom-t}
\end{eqnarray}

In order to solve Eq. (\ref{eq:eom-t}), the equation requires suitable
boundary conditions and initial condition. We choose the following
boundary conditions at the center and edge
\begin{eqnarray}
\partial_{t}\partial_{r}\chi|_{r=0}= & \partial_{t}\partial_{r}\chi|_{r=R}= & 0.
\end{eqnarray}
This means that the smoothness of the bubbles is ensured at the center
and the edge. Of course, it is also possible to choose $\partial_{t}\chi|_{r=R}=0$
at the edge, i. e., a false vacuum at the far distance all the time.
We find that there is almost no difference between the two choices
when the radial dimensions are large enough. It can be foreseen that
the choice of this paper is more reasonable when the space contains
more than one bubble. At the conformal boundary, we have
\begin{eqnarray}
\partial_{t}\partial_{z}\chi|_{z=0} & = & 0.
\end{eqnarray}
This corresponds to the fact that the current mass is fixed and its
does not vary with time.

For the initial condition, we consider adding a perturbation $\delta\chi$
to the bounce solution of the scalar field $\chi_{b}$. Because the
meaning of the bounce solution is the critical bubble solution, i.e.,
any bubble larger than it will expand, and smaller than it will shrink.
Since the bounce solution $\chi_{b}$ does not vary with time, the
selection of the perturbation is important, which will determine whether
the bubble expands or shrinks. Therefore, the perturbation $\delta\chi$
is not completely random, but is controlled to be positive or negative. 
In this paper, we choose the perturbation as follows
\begin{eqnarray}
\delta \chi(z,r)=\mathcal{A}\,\chi_t(z)\big\{\exp[(r-r_0)^2] + 1\big\}^{-1},
\end{eqnarray}
where $\mathcal{A}$ is a constant that is small enough, $\chi_t$ is the true vacuum solution at that temperature, and $r_0$ is an arbitrary number.

\begin{figure}
\includegraphics[width=0.45\textwidth]{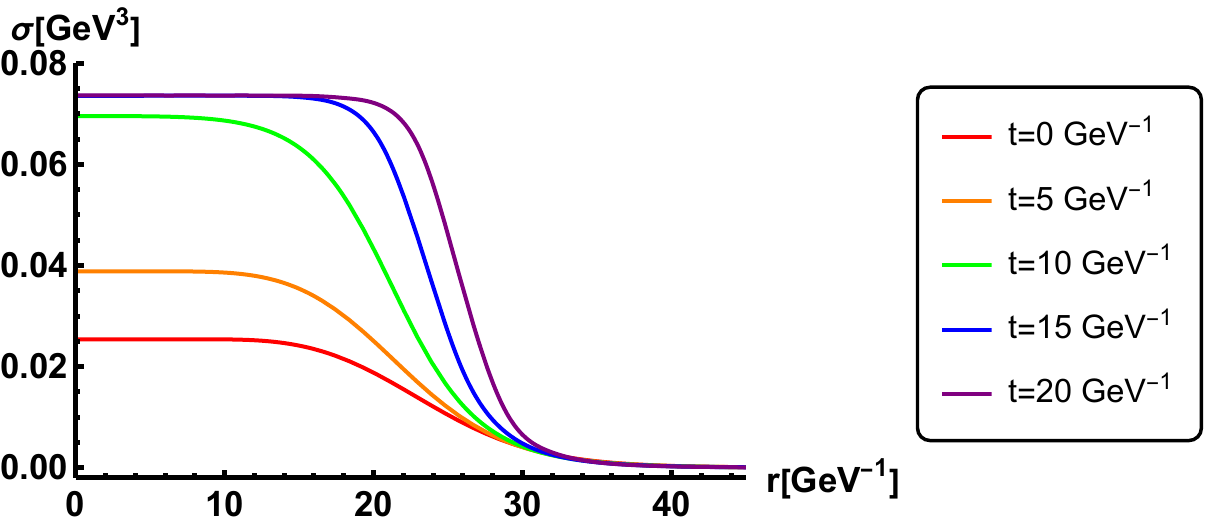}\hspace{1.5cm}\includegraphics[width=0.45\textwidth]{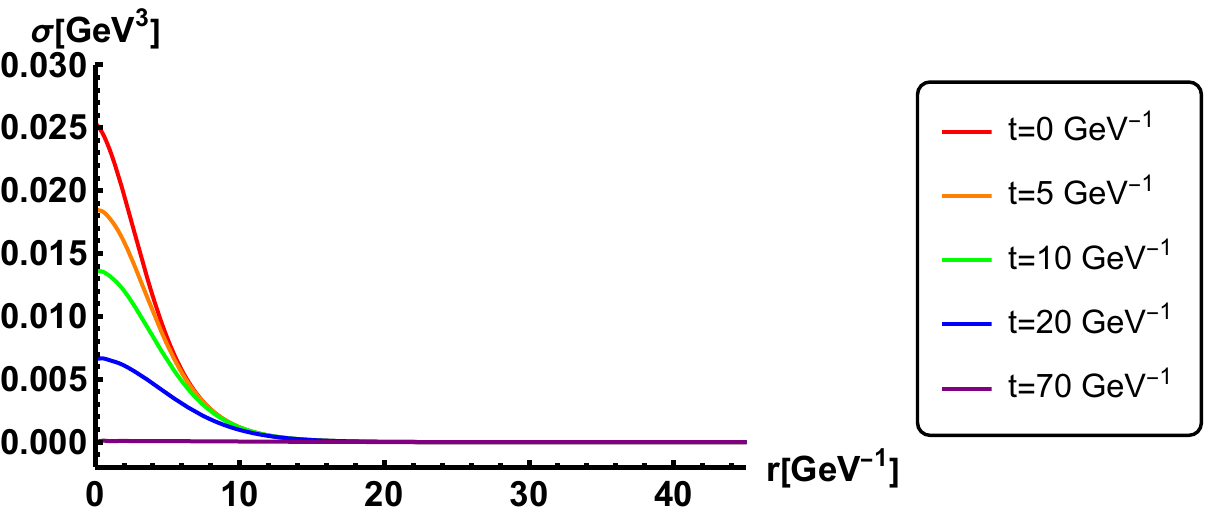}
\vskip -0.05cm \hskip 0 cm
\hskip -3 cm \textbf{( a ) } \hskip 9 cm \textbf{( b )}
\caption{\label{fig:bubble_t} The panels (a) and (b) show the profiles of
the bubbles at different times with $T=160$ MeV when positive and
negative perturbations are added, respectively.}
\end{figure}

The panels (a) and (b) of Fig. \ref{fig:bubble_t} show the evolution
of the bubble with time at $T=160$ MeV when positive and negative
perturbations are added, respectively. As seen in panel (a), when
positive perturbation is added, the bubble goes through two steps.
First the condensation value at the center of the bubble keeps
growing until the true vacuum. After that, the structure of the bubble
wall stabilizes and gradually expands outward. At times less than
about $t\simeq15{\rm GeV^{-1}}$, the bubble is still in the first
step, and when the time is greater than 15 ${\rm GeV^{-1}}$ the bubble
enters the second step. If the temperature is $170{\rm MeV}\lesssim T\lesssim174{\rm MeV}$,
then the first step of bubble expansion does not occur. Of course,
the exact timing of the distinction between the two steps is related
to the amplitude of the perturbation, which is roughly equal in magnitude
when the perturbation is small enough. The panel (b) displays the
bubble shrinking with negative perturbation. As can be seen in the
panel, the condensation value at the center of the bubble gradually
decreases until the system completely returns to the false vacuum
at $t\simeq70{\rm GeV^{-1}}$. The time of the bubble shrinking process
is roughly comparable to the time of the first step of expansion,
for bubbles that are neither too large nor too small.

\begin{figure}
\includegraphics[width=0.45\textwidth]{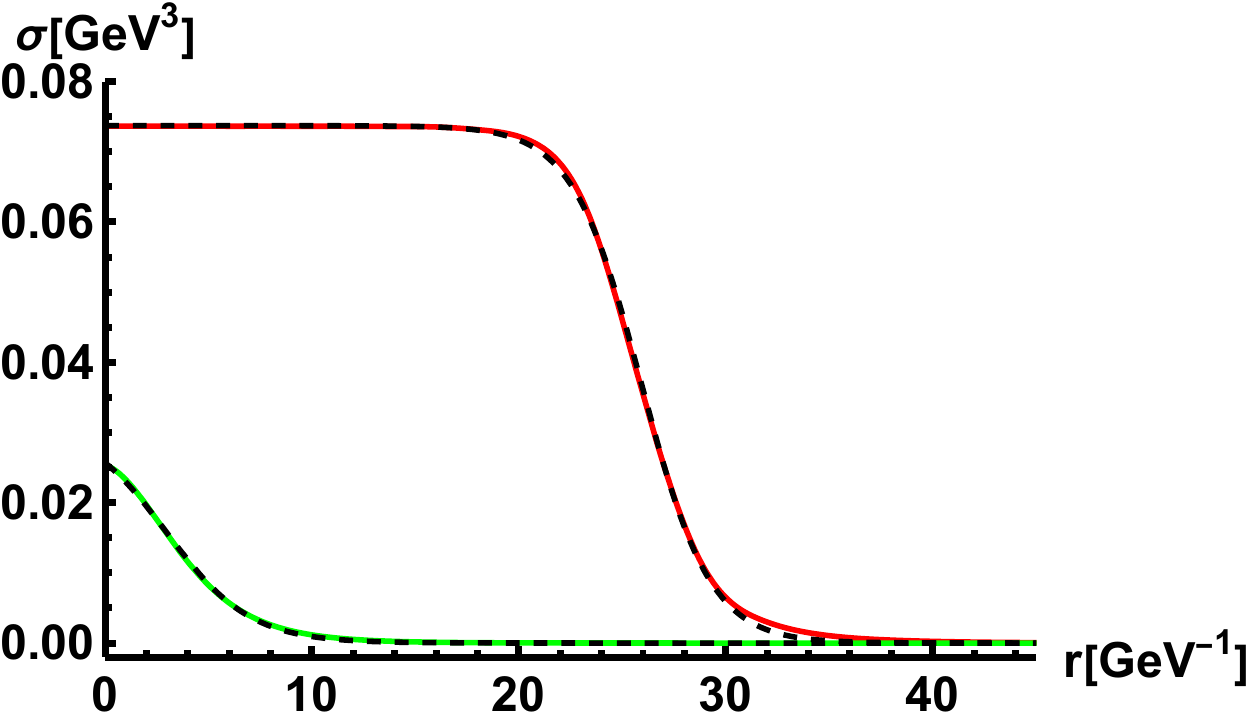}

\caption{\label{fig:fit} The condensate as a function of radial r, where the solid line is the numerical solution and the dashed line represents the fitting results of the hyperbolic tangent function.}
\end{figure}

\begin{figure}
\includegraphics[width=0.41\textwidth]{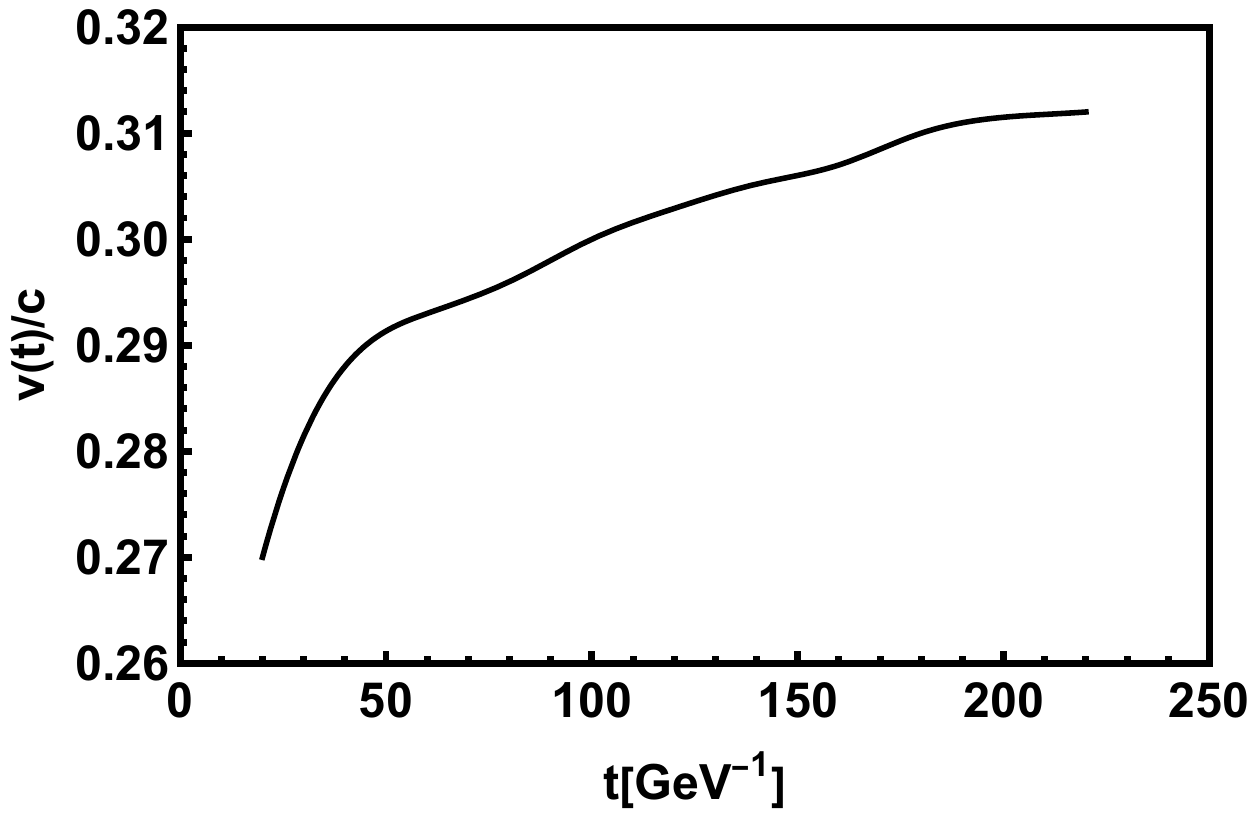}\hspace{1.5cm}\includegraphics[width=0.39\textwidth]{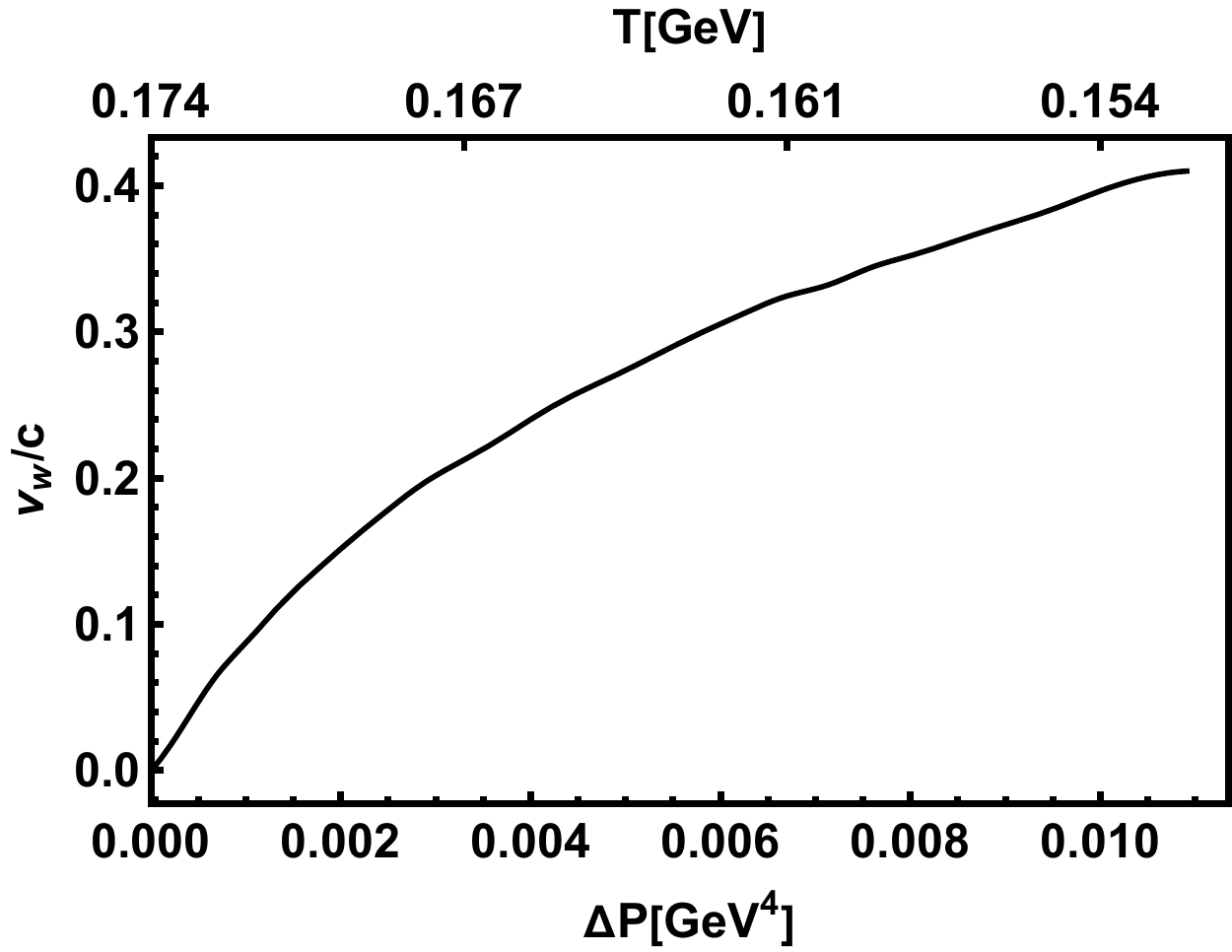}
\vskip -0.05cm \hskip 0 cm
\hskip 0.5 cm \textbf{( a ) } \hskip 7.5 cm \textbf{( b )}
\caption{\label{fig:bubble_v} The panel (a) shows the bubble wall velocity
as a function of time. The panel (b) shows the final velocity of the
bubble wall as a function of the pressure difference $\Delta P$ between
the inside and outside or the temperature $T$.}
\end{figure}

To obtain the bubble wall velocity, we used the function $\sigma(r)=\frac{\sigma_{0}(t)}{2}[-\tanh(\frac{r(t)-r_{w}(t)}{l_{w}(t)})+1]$
to fit the profile of the condensation. Here, $\sigma_{0}$ is the
condensation value at the center, $r_{w}$ denotes the radius of the
bubble, and $l_{w}$ represents the thickness of the bubble wall. Fig. \ref{fig:fit} shows the numerical and fitting results. It can be seen that the function fits the numerical results quite well except at the edges of the bubble walls.
Therefore, we can define the velocity of the bubble wall by $v(t)=dr_{w}(t)/dt$.
The panel (a) of Fig. \ref{fig:bubble_v} exhibits the bubble wall
velocity as a function of time at a temperature of 160 MeV. It can
be seen that the velocity gradually increases with time, while the
acceleration gradually decreases. The velocity reaches its final velocity
about $0.31\, c$ when the time reaches about $t\simeq200{\rm GeV^{-1}}$.
As for the relaxation time for the bubble to reach its final velocity,
it depends on the strength of the perturbation. Numerical calculations show that the relaxation times with different perturbations are roughly equivalent.
Note that the velocity is not well defined at time $t\lesssim20{\rm GeV^{-1}}$.
This is because the bubble is still in the first step during this
time and the wall configuration is changing and therefore not shown
in the panel.

The final velocity of the bubble wall $v_w$ is defined as $v_w=\lim_{t\to\infty} v(t)$. The panel (b) of Fig. \ref{fig:bubble_v} shows the final velocity
of the bubble wall as a function of the pressure difference $\Delta P$
between the inside and outside or the temperature $T$. According to the holographic principle, the pressure can be obtained from the free energy $F=-P$, as in Refs. \citep{Chelabi:2015cwn,Chelabi:2015gpc}. At low temperatures,
the final velocity is approximately linear with respect to the pressure
difference, while at temperatures close to the critical temperature,
their relation exhibits nonlinear behavior. In Ref. \citep{Janik2022a},
the relation between velocity and pressure difference has a nonlinear
behavior. Although, the method used in Ref. \citep{Janik2022a} is
different from the one used in this paper, the behavior exhibited
is very similar. It can be seen from the panel that even if the temperature
drops to 154 MeV, the final velocity still does not exceed sound speed
$c_{s}=1/\sqrt{3}$ of the plasma, which corresponds to the deflagration
case. This is in agreement with the results of Refs. \citep{Bea2021a,Bea:2021zol,Bea2022,Janik2022a}.
It is reasonable to speculate that in the bottom-up holographic model,
the final velocity is always limited to the deflagrations region because
the system contains a holographic dissipation mechanism \citep{Adams:2012pj}.
While in Ref. \citep{Bigazzi2021}, the Dp brane system can achieve
bubble wall velocities close to the speed of light. Whether detonation
and hybrid cases can be implemented in the bottom-up holographic model
is still an open question.

\section{bubbles and thin wall approximation}
\label{sec:4}

In the previous section, we have obtained the bounce solution. By
analyzing the properties of the holographic bounce solution, the characteristics
of the strong FOPT can be understood qualitatively,
which helps to understand its GW power spectrum. We know that there
are three important characteristic temperatures during the dynamical
FOPT that exhibit the details of the phase
transition and are finally reflected in the GW. The first one is the
critical temperature $T_{c}$ of the phase transition, i.e., the temperature
at which the free energy has a degenerate minimum. The rest are the nucleation temperature
$T_{n}$ and percolation temperature $T_{p}$, which represent the
temperature of beginning nucleation and ending of phase transition,
respectively. The temperature $T_{n}$ and $T_{p}$ are mainly determined
by the profile of the Euclidean action $\frac{S_{b}}{T}$ of bounce
solution with temperature. In addition, the bubble wall thickness
and bubble radius can be obtained from the bounce solution. In this
section, we analyze the holographic behavior of the bounce solution
with temperature. It should be noted that we only show the QCDPT with
$\upsilon_{3}\neq0$ (Model I). For $\upsilon_{6}\neq0$ (Model II) or EWPT,
the numerical results change slightly and the conclusions in the following
still apply.

\begin{figure}
\includegraphics[width=0.45\textwidth]{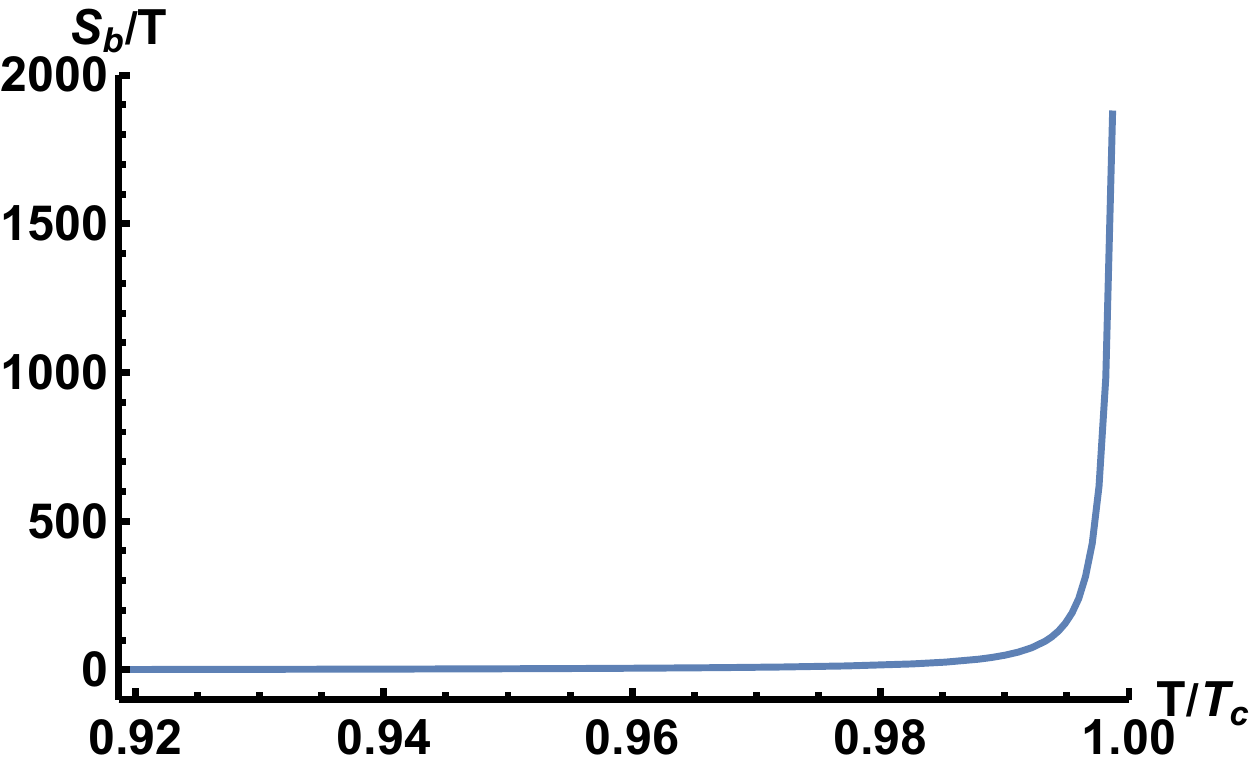}

\caption{\label{fig:action} The Euclidean action $\frac{S_{b}}{T}$
as a function of temperature.}
\end{figure}

From Eq. (\ref{eq:onshellaction}) we can calculate the Euclidean
action $\frac{S_{b}}{T}$. Fig. \ref{fig:action} shows the on-shell
action $\frac{S_{b}}{T}$ as a function of $\frac{T}{T_{c}}$ calculated
by the bounce solution. It can be seen from the figure that the value
of the action decreases rapidly as the temperature decreases, and
at about $\frac{T}{T_{c}}\sim0.99$, i.e., at a temperature of about
172 MeV, $\frac{S_{b}}{T}$ drops to 30. The nucleation temperature
$T_{n}$ can be given by relation $\frac{S_{b}(T_{n})}{T_{n}}\sim180$
for QCDPT ($\frac{S_{b}(T_{n})}{T_{n}}\sim140$ for EWPT) (a more
rigorous definition is shown in the next section). Therefore, the
nucleation temperature is very close to the critical temperature $T_{c}\simeq T_{n}$.
Furthermore, at low temperatures, the action does not have a local
minimum. Also, since the nucleation probability
$\Gamma$ is proportional to the action, it can be seen that bubbles
are created rapidly and abundantly as the temperature decreases. Consequently,
it is reasonable to suppose that the percolation temperature is close
to the nucleation temperature $T_{n}\simeq T_{p}$. As shown above,
the three characteristic temperatures have relation $T_{c}\simeq T_{n}\simeq T_{p}$,
which corresponds to the supercooling case \citep{Eichhorn:2020upj}.

\begin{figure}
\includegraphics[width=0.4\textwidth]{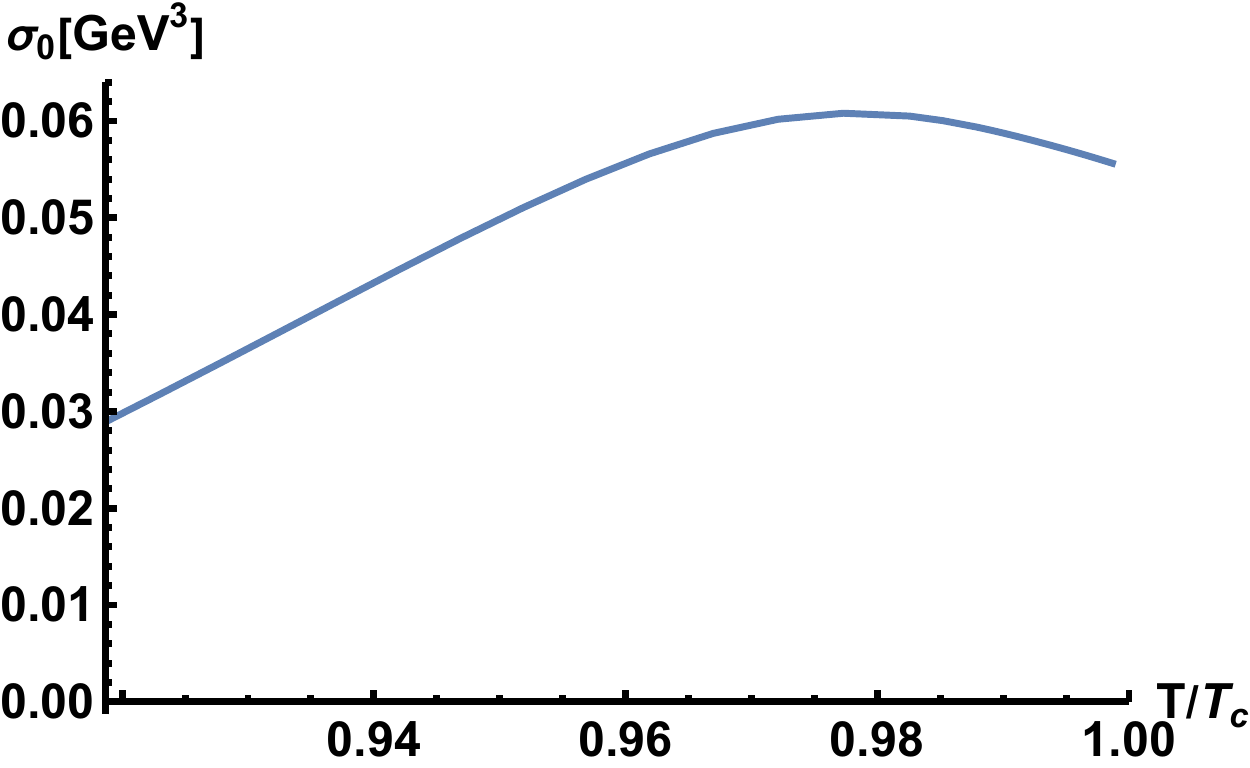}\hphantom{}
\vskip -0.05cm \hskip 0 cm
\textbf{( a ) }
\vskip 0.5cm
\includegraphics[width=0.4\textwidth]{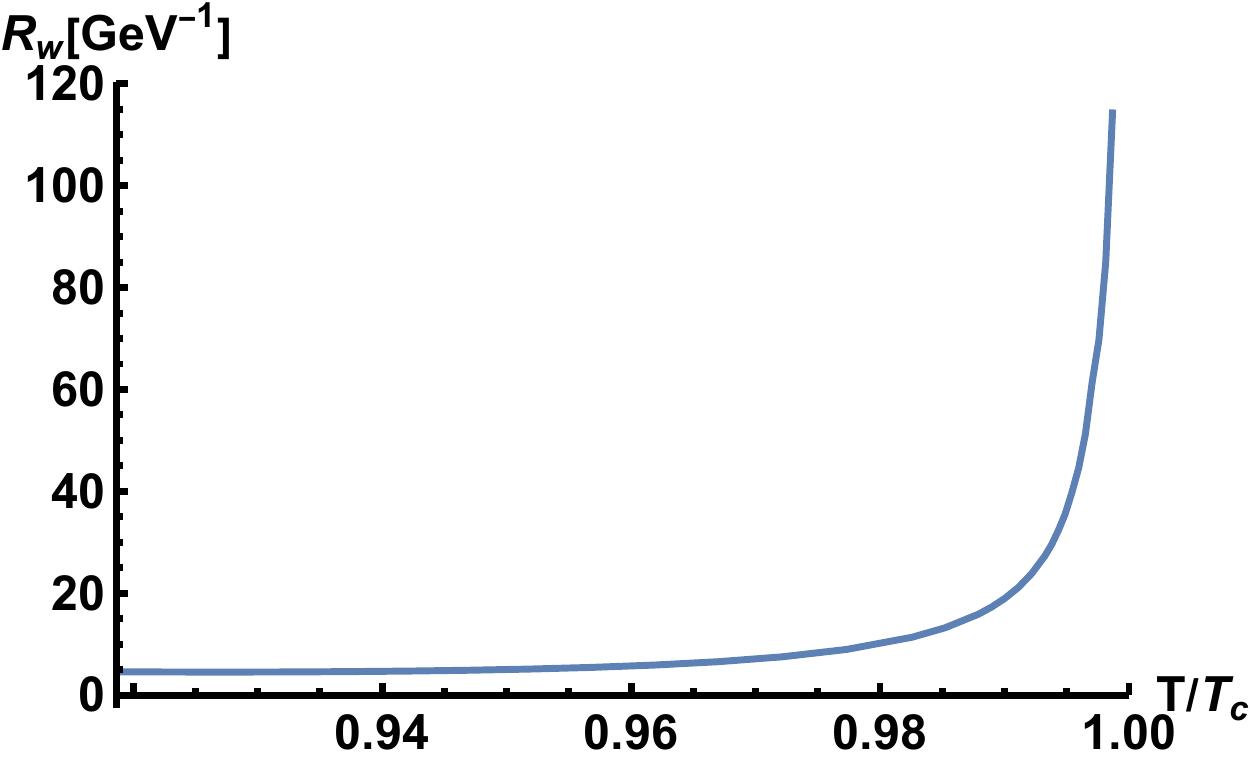}\hspace{1.5cm}\includegraphics[width=0.4\textwidth]{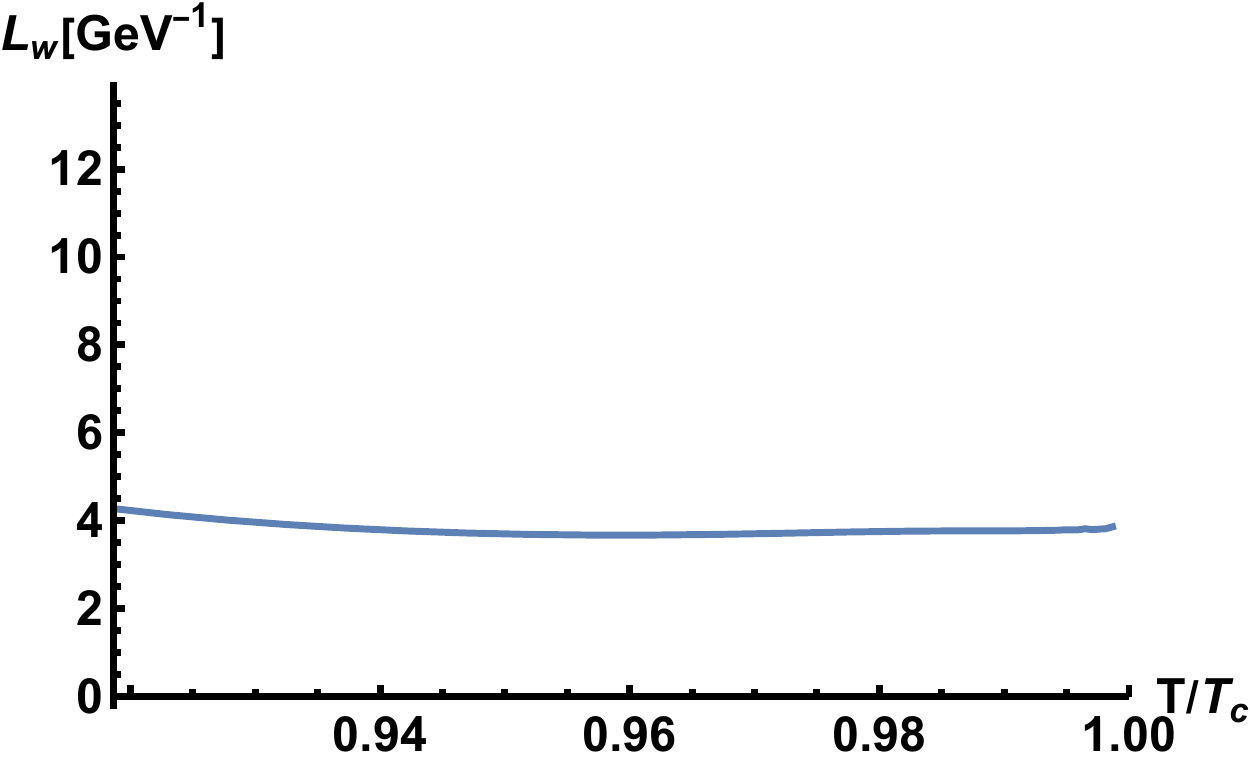}
\vskip -0.05cm \hskip 0 cm
\hskip 0.5 cm \textbf{( b ) } \hskip 7.5 cm \textbf{( c )}
\caption{\label{fig:bubble_tanh}The panels (a), (b) and (c) show the condensation
value at the center $\sigma_{0}$, the bubble radius $R_{w}$ and
the bubble wall thickness $L_{w}$ as functions of the temperature,
respectively.}
\end{figure}

As in the previous section, we use hyperbolic tangent function interpolation
condensation as a function of the radial $r$, shown following
\begin{eqnarray}
\sigma(T,r) & = & \frac{\sigma_{0}(T)}{2}[-\tanh(\frac{r-R_{w}(T)}{L_{w}(T)})+1],
\end{eqnarray}
where $\sigma_{0}$ is the condensation value at the center, $R_{w}$
denotes the radius of the bubble, and $L_{w}$ represents the thickness
of the bubble wall. Unlike the previous section, these quantities
are independent of time since the critical bubbles do not evolve with
time. Fig. \ref{fig:bubble_tanh} displays the condensation $\sigma_{0}$,
radius $R_{w}$, and thickness $L_{w}$ as functions of temperature.
As seen in panel (a), the condensation at the center increases with
increasing temperature at $T\lesssim0.975T_{c}\sim170$MeV and has
the opposite behavior at $T\gtrsim0.975T_{c}$. Two factors with opposite
effects influence this behavior. As can be seen in Fig. \ref{fig:sigma_T},
the condensation value decreases continuously with increasing temperature;
in contrast, as in Fig. \ref{fig:bubble_T}, the critical bubble is
larger at higher temperatures, which favors the generation of a true
vacuum at the center. For the bubble radius $R_{w}$, panel (b) shows
that the critical radius varies slightly at temperature $T\lesssim0.98T_{c}\sim171$MeV.
However, at temperature $T\gtrsim0.98T_{c}$, the critical radius
of the bubble increases rapidly, which is not favorable for bubble
generation. In contrast to the behavior of the bubble radius, the
bubble wall thickness essentially does not vary with temperature.
And the thickness changes slightly at low temperature $T\lesssim0.94T_{c}\sim164$MeV. 
This is because the hyperbolic function fitting is not a good choice
at this time.

\subsection{Thin-wall approximation}

From the previous discussion, it is clear that in this holographic
model, the FOPT is supercooled. As
in Ref. \citep{Eichhorn:2020upj}, the thin-wall approximation can
be applied in this case. Under this approximation, many physical quantities,
such as latent heat, duration time of phase transition and surface
tension of bubble, have simpler forms and can be obtained more easily.
In this section, we estimate the surface tension and the parameters
$\alpha$ and $\beta/H$ using the Euclidean action obtained previously.
Among them, the parameters $\alpha$ and $\beta/H$ are very important
for the GW power spectrum.

\begin{figure}
\includegraphics[width=0.45\textwidth]{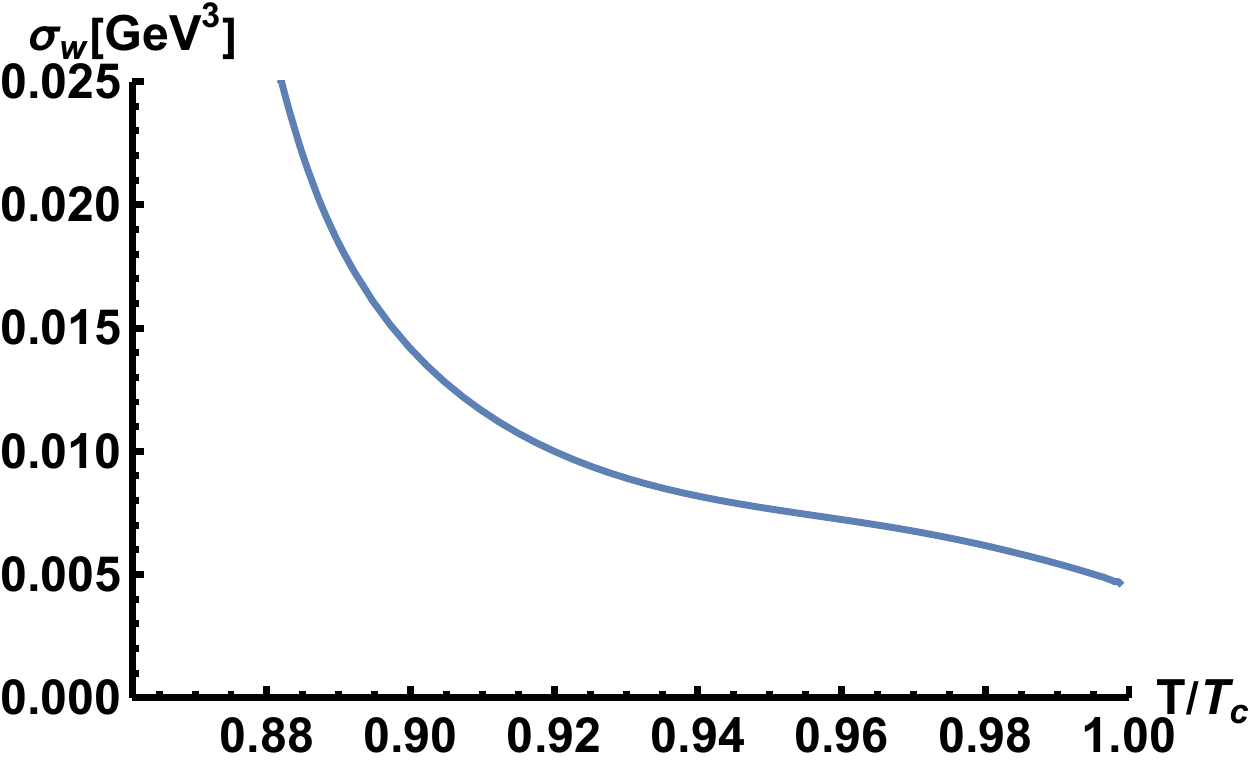}

\caption{\label{fig:bubble_tension} The surface tension
$\sigma_{w}$ of the bubble wall as a function of temperature.}
\end{figure}

From Ref. \citep{Eichhorn:2020upj}, the parameter $\alpha$ is

\begin{eqnarray}
\alpha & \simeq & \frac{15}{2\pi^{2}}\frac{L_{c}}{g_{*}(T_{c})T_{c}^{4}},
\end{eqnarray}
where $g_{*}$ is the relativistic degree of freedom of the system
and $L_{c}$ is the latent heat of phase transition. For QCDPT (EWPT),
the degrees of freedom can be approximated as $g_{*}\simeq10$ ($g_{*}\simeq100$).
Under the thin-wall approximation, the latent heat has the following
form

\begin{eqnarray}
L_{c} & = & -T\frac{\partial \Delta V_{\text{eff}}(\langle\phi\rangle_{T},T)}{\partial T}\bigg|_{T=T_{c}}=T\frac{\partial \Delta F(\langle\phi\rangle_{T},T)}{\partial T}\bigg|_{T=T_{c}},
\end{eqnarray}
with difference of the free energy $\Delta F$. The calculation of the free energy $F$
can be found in Refs. \citep{Chelabi:2015cwn,Chelabi:2015gpc}. Within
the approximation, the three-dimensional Euclidean action $S_{b}$
can be divided into two parts, as follows
\begin{eqnarray}
S_{b}(T) & = & -\frac{4\pi}{3}R_{w}(T)\varepsilon(T)+4\pi R_{w}(T)^{2}\sigma_{w}(T),
\end{eqnarray}
with the $\varepsilon(T)=\Delta  V_{\text{eff}}(\langle\phi\rangle_{T},T)$ and surface tension $\sigma_{w}$
of the bubble wall. From the above equation, it can be seen that the
energy of the critical bubble consists of the vacuum energy inside
the bubble and the surface energy of the bubble wall. With the bubble
radius previously calculated, the surface tension can be obtained
as shown in Fig. \ref{fig:bubble_tension}. For the low-temperature
region in the figure, the result is not reliable because the thin-wall
approximation no longer applies.

Also from Ref. \citep{Eichhorn:2020upj}, the inverse of duration time $\beta/H$
of the phase transition is given as
\begin{eqnarray}
\frac{\beta}{H} & = & \left(\frac{3}{4\pi}\frac{T_{c}L_{c}^{2}}{\sigma_{w}(T_{c})^{3}}\right)^{1/2}\left(\frac{S_{b}(T_{p})}{T_{p}}\right)^{3/2},
\end{eqnarray}
with the percolation temperature $T_{p}$. With the above definitions
and the relation $T_{c}\simeq T_{n}\simeq T_{p}$, the final results
of parameters $\alpha$ and $\beta/H$ are shown in Tab. \ref{tab:paras_thin}.
Here, the velocity $v_w$ is calculated at the percolation temperature $T_p$.
From the table, we can find that for QCDPT, the parameter $\alpha$
is greater than 1, about 4-5, which is strongly FOPT case, and the parameter $\beta/H$ is large, which means
that the phase transition ends rapidly. For EWPT, the strength parameter $\alpha$ is less than 1, while the inverse of the duration time remains large, which means that the phase transition is sufficiently fast, that is, the so-called weakly supercooled FOPT.

\begin{table}
\begin{tabular}{|c|c|c|}
\hline 
Thin-wall approx & QCDPT & EWPT \tabularnewline
\hline 
\hline 
$\alpha$ & 4-6 & 0.4-0.6\tabularnewline
\hline 
$\beta/H$ & 30000-60000 & 6000-20000\tabularnewline
\hline 
$v_{w}$ & 0.04 & 0.1\tabularnewline
\hline 
\end{tabular}

\caption{\label{tab:paras_thin} The parameters $\alpha$,
$\beta/H$ and the bubble wall velocity $v_\omega$ estimated by the thin-wall approximation during QCDPT and
EWPT.}

\end{table}

\section{stochastic gravitational waves}
\label{sec:5}

When the system reaches the nucleation temperature, bubbles will be created, and they will
collide and merge together in the plasma. Then the false vacuum will be transformed into
the true vacuum. In this process, part of the system's energy is eventually
converted into GW radiation. Within the linear approximation,
the total GW power spectra can be written as
\begin{eqnarray}
h^{2}\Omega_{{\rm GW}} & \simeq & h^{2}\Omega_{{\rm coll}}+h^{2}\Omega_{{\rm sw}}+h^{2}\Omega_{{\rm turb}},
\end{eqnarray}
where $\Omega_{{\rm coll}}$ comes from bubble collisions \citep{Kosowsky:1991ua,Kosowsky:1992rz,Kosowsky:1992vn,Kamionkowski:1993fg,Caprini:2007xq,Huber:2008hg,Caprini:2009fx,Espinosa:2010hh,Weir:2016tov,Jinno:2016vai},
$\Omega_{{\rm sw}}$ from acoustic waves in the plasma after the collision
\citep{Hindmarsh:2013xza,Giblin:2013kea,Giblin:2014qia,Hindmarsh:2015qta,Hindmarsh:2017gnf},
and $\Omega_{{\rm turb}}$ from magnetohydrodynamic turbulence in
the plasma \citep{Kosowsky:2001xp,Dolgov:2002ra,Caprini:2006jb,Gogoberidze:2007an,Kahniashvili:2008pe,Kahniashvili:2009mf,Caprini:2009yp,Kisslinger:2015hua}. 

In the previous section we have found that for the holographic model,
the final velocity of the bubble wall is less than the speed of sound,
i.e., the non-runaway case. According to Refs. \citep{Kamionkowski:1993fg,Espinosa:2010hh,Caprini:2015zlo,Ellis:2019oqb,Alanne:2019bsm,Cutting:2019zws},
the GWs generated by collisions in the non-runaway
case can be neglected with respect to acoustic waves and magnetohydrodynamic
turbulence. Therefore, the total power spectrum is approximated to
\begin{eqnarray}
h^{2}\Omega_{{\rm GW}} & \simeq & h^{2}\Omega_{{\rm sw}}+h^{2}\Omega_{{\rm turb}}.
\end{eqnarray}
From numerical simulations \citep{Hindmarsh:2015qta}, the power spectrum
from sound waves has the form of
\begin{eqnarray}
h^{2}\Omega_{{\rm sw}}(f) & = & 2.65\times10^{-6}\left(\frac{H}{\beta}\right)\left(\frac{\kappa_{v}\alpha}{1+\alpha}\right)^{2}\left(\frac{100}{g_{*}}\right)^{\frac{1}{3}}v_{w}S_{{\rm sw}}(f),\\
S_{{\rm sw}}(f) & = & (f/f_{{\rm sw}})^{3}\left(\frac{7}{4+3(f/f_{{\rm sw}})^{2}}\right)^{7/2},\\
f_{{\rm sw}} & = & 1.9\times10^{-2}{\rm mHz}\frac{1}{v_{w}}\left(\frac{\beta}{H}\right)\left(\frac{T_{p}}{100{\rm GeV}}\right)\left(\frac{g_{*}}{100}\right)^{\frac{1}{6}},
\end{eqnarray}
where the factor $\kappa_{v}$ represents the ratio of vacuum energy
transformed into bulk motion. The specific form of the factor $\kappa_{v}$
depends on the bubble wall velocity and has the following form in
the limits of large and small velocities
\begin{eqnarray}
\kappa_{v} & \simeq & \begin{cases}
\alpha\left(0.73+0.083\sqrt{\alpha}+\alpha\right)^{-1}, & v_{w}\sim1,\\
v_{w}^{6/5}6.9\alpha\left(1.36-0.037\sqrt{\alpha}+\alpha\right)^{-1}, & v_{w}\lesssim0.1.
\end{cases}
\end{eqnarray}
From the Tab. \ref{tab:paras_thin}, it is obtained that
the velocity is about $v_{w}\sim0.1$, since the phase transition
of the holographic model is supercooled. Therefore, in this
paper, the expression for the factor $\kappa_{v}$ is
\begin{eqnarray}
\kappa_{v} & = & v_{w}^{6/5}6.9\alpha\left(1.36-0.037\sqrt{\alpha}+\alpha\right)^{-1}.
\end{eqnarray}

For the power spectrum from the Kolmogorov-type turbulence, numerical
simulations show that it can be given as \citep{Kosowsky:2001xp,Caprini:2009yp,Binetruy:2012ze}
\begin{eqnarray}
h^{2}\Omega_{{\rm turb}}(f) & = & 3.35\times10^{-4}\left(\frac{H}{\beta}\right)\left(\frac{\kappa_{{\rm turb}}\alpha}{1+\alpha}\right)^{\frac{3}{2}}\left(\frac{100}{g_{*}}\right)^{1/3}v_{w}S_{{\rm turb}}(f),\\
S_{{\rm turb}}(f) & = & \frac{(f/f_{{\rm turb}})^{3}}{\left[1+(f/f_{{\rm turb}})\right]^{\frac{11}{3}}\left(1+8\pi f/h_{*}\right)},\\
f_{{\rm turb}} & = & 2.7\times10^{-2}{\rm mHz}\frac{1}{v_{w}}\left(\frac{\beta}{H}\right)\left(\frac{T_{p}}{100{\rm GeV}}\right)\left(\frac{g_{*}}{100}\right)^{\frac{1}{6}},
\end{eqnarray}
where the Hubble rate $h_{*}$ is
\begin{eqnarray}
h_{*} & = & 16.5\times10^{-3}{\rm mHz}\left(\frac{T_{p}}{100{\rm GeV}}\right)\left(\frac{g_{*}}{100}\right)^{\frac{1}{6}}.
\end{eqnarray}
In this paper, referring to the numerical results of the Refs. \citep{Hindmarsh:2015qta,Caprini:2015zlo},
the factor $\kappa_{{\rm turb}}$ is chosen to be
\begin{eqnarray}
\kappa_{{\rm turb}} & = & 0.05\kappa_{v}.
\end{eqnarray}
It should be noted that, due to the complexity of turbulence, more
research is still needed on the exact form of the factor $\kappa_{{\rm turb}}$
and its relation to $\kappa_{v}$. In this holographic model, the
choice can be reasonable considering that the phase transition may
end rapidly and therefore the effect of turbulence is expected to
be suppressed.

In the GW power spectrum, the parameters $\alpha$ and $\beta/H$
are important, which represent the vacuum energy release and duration
time of the phase transition, respectively. The parameter $\alpha$
is defined as
\begin{eqnarray}
\alpha & \equiv & \frac{1}{\rho_{\text{rad}}}\left(\Delta V_{\text{eff}}-\frac{T}{4}\frac{\partial\Delta V_{\text{eff}}}{\partial T}\right)\bigg|_{T=T_{p}}=-\frac{1}{\rho_{\text{rad}}}\left(\Delta F-\frac{T}{4}\frac{\partial\Delta F}{\partial T}\right)\bigg|_{T=T_{p}},
\end{eqnarray}
with radiation energy density
\begin{eqnarray}
\rho_{\text{rad}} & = & g_{*}\frac{\pi^{2}}{30}T^{4}.
\end{eqnarray}
The inverse of the duration time is defined as
\begin{eqnarray}
\frac{\beta}{H} & \equiv & T\frac{d}{dT}\left(\frac{S_{b}}{T}\right)\bigg|_{T=T_{p}}.
\end{eqnarray}

The definitions of the other two characteristic temperatures of the
phase transition, i. e., nucleation temperature $T_{n}$ and percolation
temperature $T_{p}$, are shown below. The nucleation temperature
is defined as one bubble per unit Hubble volume and is written as

\begin{eqnarray}
N(T_{n}) & = & \int_{T_{n}}^{T_{c}}\frac{\mathrm{d}T}{T}\frac{\Gamma(T)}{H(T)^{4}}=1,
\end{eqnarray}
where the nucleation rate $\Gamma(T)$ is given in Eq. (\ref{eq:bubble_p})
and the Hubble parameter $H(T)$ is

\begin{eqnarray}
H(T) & = & \sqrt{\frac{\rho_{\text{rad}}+\rho_{\text{vac}}}{3M_{\text{pl}}^{2}}},
\end{eqnarray}
with reduced Planck mass $M_{\text{pl}}=2.435\cdot10^{18}$ GeV. Referring
to Refs. \citep{Guth:1979bh,Guth:1981uk,Rintoul_1997}, the probability
of a false vacuum is defined as
\begin{eqnarray}
P(T) & = & e^{-I(T)},
\end{eqnarray}
with
\begin{eqnarray}
I(T) & = & \frac{4\pi}{3}\int_{T}^{T_{c}}\mathrm{d}T'\frac{\Gamma(T')}{H(T')T'^{4}}\left(\int_{T}^{T'}\mathrm{d}T''\frac{v_{w}(T'')}{H(T'')}\right)^{3}.
\end{eqnarray}
The percolation temperature is defined as $I(T_{p})\simeq0.34$, which
is the temperature when the probability of false vacuum is about $P(T_{p})\simeq0.7$.

The Tab. \ref{tab:result} shows the quantities related to the phase
transition in different holographic models. 
Here, the velocity $v_w$ is selected at the percolation temperature $T_p$.
Compare to Table I, we
can see that the rigorous results are very close to that of thin-wall
approximation. This further verifies the reliability of the approximation
in the holographic model. It should be noted that the strength parameter
$\alpha$ of QCDPT is greater than 1, which would be a strongly supercooled
FOPT. For EWPT, FOPT is weakly supercooled. It is worth noting that although QCDPT is strongly supercooled, its properties are very similar to those of the EWPT and there are no minimum values of the weights $S_b/T$ as mentioned in the Ref. \citep{Eichhorn:2020upj}.
Furthermore, for parameters $\alpha$ and $\beta/H$, we find an inverse
relation between them, which is similar to that of Ref. \citep{Eichhorn:2020upj}.
For Models I and II, we found that Model II $(\upsilon_{6}\neq0)$
has lower percolation temperature and larger bubble wall velocity,
which favors a bigger GW signal.

From Tab. \ref{tab:result}, we also find that the results of the holographic model differ significantly from those of the traditional calculations in quantum field theory. In quantum field theory, the parameter $\alpha$ is generally not larger than 0.01 and the latent heat released in the phase transition is not large. However, the strength of the phase transition is greater than 0.1 for both color brane \citep{Chen:2017cyc} and flavor brane in the holographic model. For the parameter $\beta/H$, field theory calculations show that it is generally not larger than 1000. But for the holographic model, the parameter $\beta/H$ is larger than 5000 during QCDPT or EWPT. This means that the duration time of the strongly coupled phase transition is shorter compared to the weakly coupled case. As for the bubble wall velocity, one would expect it to be close to the speed of light $c$. Unfortunately, the holographic results suggest that the bubble wall velocity is smaller than the sound speed of the plasma.

\begin{table}
\begin{tabular}{|c|c|c|c|c|}
\hline 
 & \multicolumn{2}{c|}{QCDPT} & \multicolumn{2}{c|}{EWPT}\tabularnewline
\hline 
Model & I $(\upsilon_{3}\neq0)$ & II $(\upsilon_{6}\neq0)$ & I $(\upsilon_{3}\neq0)$ & II $(\upsilon_{6}\neq0)$\tabularnewline
\hline 
\hline 
$g_{*}$ & \multicolumn{2}{c|}{10} & \multicolumn{2}{c|}{100}\tabularnewline
\hline 
$\alpha$ & 4.881 & 6.142 & 0.238 & 0.763\tabularnewline
\hline 
$\beta/H$ & 41151 & 23276 & 17198 & 7238\tabularnewline
\hline 
$v_{w}$ & 0.027 & 0.041 & 0.063 & 0.125\tabularnewline
\hline 
$T_{c}${[}GeV{]} & \multicolumn{2}{c|}{0.1741} & \multicolumn{2}{c|}{122.1}\tabularnewline
\hline 
$T_{n}${[}GeV{]} & 0.1733 & 0.1712 & 120.7 & 118.1\tabularnewline
\hline 
$T_{p}${[}GeV{]} & 0.1732 & 0.1703 & 120.4 & 117.6\tabularnewline
\hline 
\end{tabular}

\caption{\label{tab:result} The critical temperature $T_{c}$, nucleation
temperature $T_{n}$, percolation temperature $T_{p}$ of phase transition,
parameters $\alpha$, $\beta/H$, and bubble wall velocity $v_{w}$
in different holographic models.}
\end{table}

\begin{figure}
\includegraphics[width=0.34\textwidth]{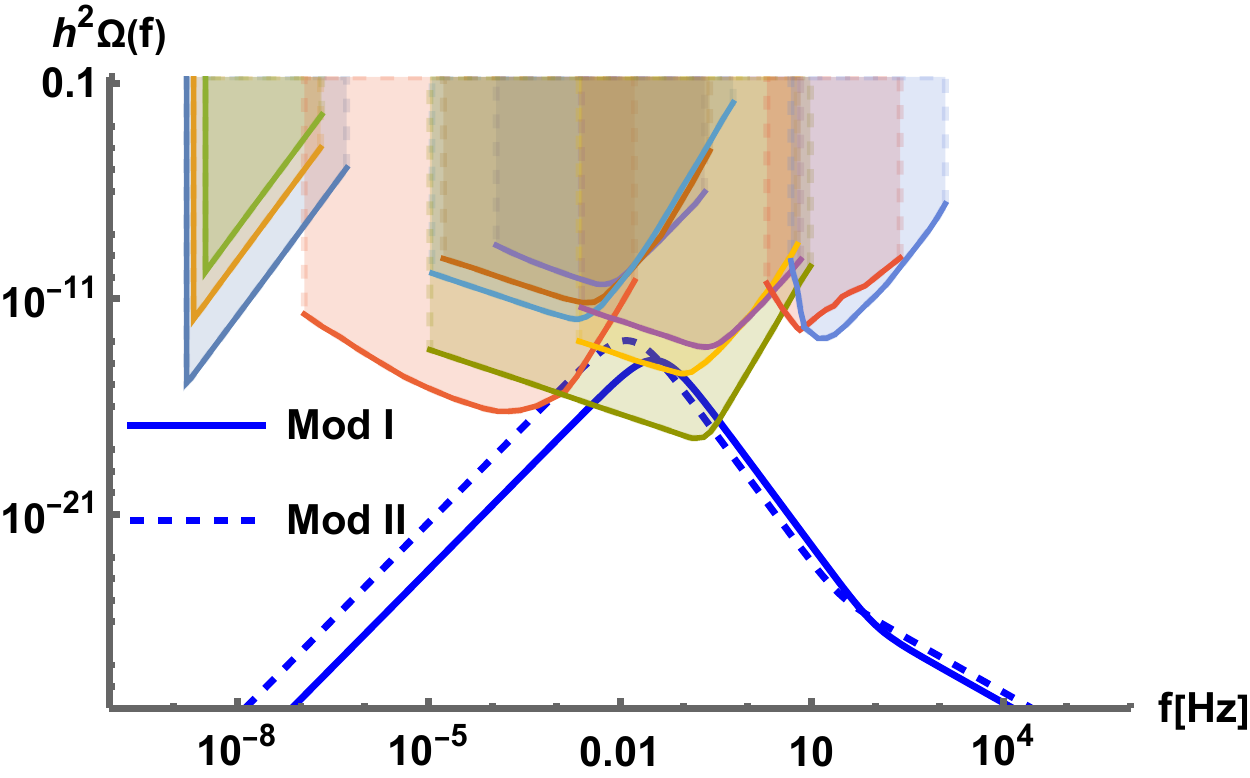}\hspace{0.5cm}\includegraphics[width=0.51\textwidth]{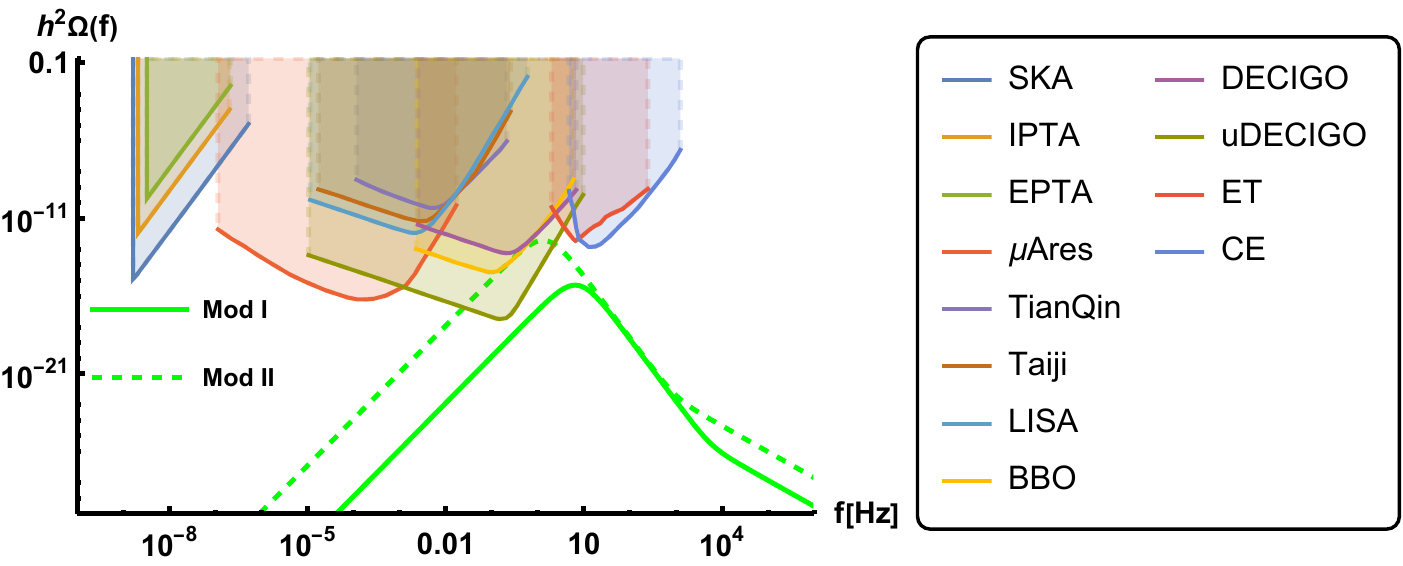}
\vskip -0.05cm \hskip 0 cm
\hskip -3 cm \textbf{( a ) } \hskip 6 cm \textbf{( b )}
\caption{\label{fig:GW} The GW power spectra in different holographic models,
where the blue (a) and green (b) lines represent QCDPT and EWPT, respectively.}

\end{figure}

The GW power spectra of QCDPT and EWPT for different holographic models
are exhibited in Fig. \ref{fig:GW}. Due to the large value of $\beta/H$,
the peak frequency of GW is larger compared to the results of quantum
field theory, causing a right shift in the sensitive frequency interval
of GW. For the holographic model of QCDPT, the peak frequency is about
0.01 Hz. In addition, since the bubble wall velocity $v_{w}$ is much
less than the speed of light $c$, the GW signal $h^{2}\Omega$ can only
reach about $10^{-13}-10^{-14}$, which can be detected by $\mu$Ares, BBO and
Ultimate-DECIGO for Model II, and by BBO and Ultimate-DECIGO for Model I. For EWPT, the peak frequency is around $1-10$ Hz,
when the GW spectrum $h^{2}\Omega$ reaches about $10^{-12}-10^{-16}$.
For Model I, the GW can be detected by BBO
and Ultimate-DECIGO, but for Model II, it is not detectable by future
experiments.

\section{conclusion and discussion}
\label{sec:6}

In this paper, holographic bounce solutions, bubble wall velocities
and GW power spectra are studied and discussed in
holographic QCD and EW models. For holographic models with first-order phase transitions, holographic bounce solutions can be obtained
by setting appropriate boundary conditions. By adding positive and
negative perturbations to the bounce solution, the bubble expands
or shrinks and the final velocity of the bubble wall can be given.
It turns out that the final velocity is still smaller than the speed
of sound in the plasma, i. e. the deflagration case, even if the phase
transition temperature is very low. Moreover, we find that the critical
temperature $T_{c}$, nucleation temperature $T_{n}$ and percolation
temperature $T_{p}$ of the phase transition are close to each other, which makes the holographic bubble
fit well with the results of thin-wall approximation. Furthermore, the strength parameter $\alpha$ calculated by the holographic model is about $\alpha\sim 5$ for QCDPT ($\alpha\sim 0.5$ for EWPT) and the inverse of the duration time $\beta/H$ is about $\beta/H\sim 10000$, and they are quite different from the weakly coupled field theory results. For QCDPT, the
GW power spectrum can reach $10^{-13}-10^{-14}$ around
the peak frequency 0.01 Hz, which can be detected by $\mu$Ares, BBO and
Ultimate-DECIGO for $\upsilon_{6}\neq0$ (Model II), and by BBO and Ultimate-DECIGO for $\upsilon_{3}\neq0$ (Model I).
For EWPT, the GW power spectrum can reach $10^{-12}-10^{-16}$
around the peak frequency $1-10$ Hz, which can be detected by BBO and Ultimate-DECIGO for $\upsilon_{6}\neq0$ (Model II), but not by future experiments for $\upsilon_{3}\neq0$ (Model I).

Referring to the quantum bounce solutions proposed by Coleman and
Callan\citep{Coleman:1977py,Callan:1977pt} with Neumann boundary
condition at the center and Dirichlet boundary condition at the edge,
the holographic bounce solution can be yielded. We found that the
radius of the critical bubble and the value of condensation at the
center reduced with decreasing temperature. If a small positive or
negative perturbation is added to the holographic bounce solution,
the bubble starts to expand or shrink. Through a long enough evolution,
the velocity of the bubble wall reaches a constant value. It turns
out that in this holographic model, the bubble velocity is deflagration,
i.e., it is less than the speed of sound of the system. Although our
calculations are under the probe approximation, that is, without a
push to the fluid, the conclusions are similar to the Refs. \citep{Bea2021a,Bigazzi2021,Bea:2021zol,Bea2022a,Bea2022,Janik2022a}.
How to get detonation and hybrid cases in the bottom-up model remains
an open question.

The radius of the critical bubble and the thickness of the bubble
wall can be obtained by the Euclidean action. We found that the bubble
radius drops rapidly with decreasing temperature while the wall thickness
varies little with temperature. In addition, we find that the critical
temperature, nucleation temperature and percolation temperature are
close to each other during the holographic phase transition. Therefore
the thin-wall approximation is considered to estimate the strength
and duration time of the phase transition. It turns out that the more
rigorous calculations agree well with the approximation results, verifying
the reliability of the approximation.

With the holographic bounce solution and the bubble final velocity,
we calculated the GW power spectra of QCDPT and EWPT. 
We find that the strength parameter $\alpha$ and the inverse of the duration time $\beta/H$ calculated by the holographic model differ significantly from the weakly coupled field theory results, which impact the GW signal. The parameter $\alpha$ can reach about 5 for QCDPT, which is strong phase transition, and about 0.5 for EWPT. The parameter $\beta/H$ is about 30,000 for QCDPT and about 10,000 for EWPT, so the holographic phase transition is sufficiently fast. 
Due to the large parameter $\beta/H$, the peak frequency shifts rightward compared to the
field theory results.
In addition, the GW spectrum is suppressed due to the small bubble
expansion speed. Eventually, for QCDPT, it can be detected by $\mu$Ares, BBO and
Ultimate-DECIGO for Model II and by BBO and Ultimate-DECIGO for Model I. For EWPT, Model II is detectable,
while Model I cannot be detected by future experiments. 
Whether it is QCDPT or EWPT, the FOPT caused by the sextic term has a larger $\alpha$ and smaller $\beta/H$ and higher GW energy than that caused by the cubic term.

For primordial black holes, the formula of mechanism \citep{Liu:2021svg}
cannot be applied because the speed of the bubble at the phase transition
is much smaller than the speed of light $c$. Therefore, it cannot
be determined whether the black holes have sufficient probability
to be produced. However, since the small bubble wall velocity may
lead to a small energy density perturbation during the collision,
we can speculate that the primordial black hole is difficult to produce
in this holographic model. Also, as in Ref. \citep{Shao:2022oqw}, the large parameter $\beta/H$ is not favorable for the formation of primordial black holes.
\begin{acknowledgments}
We thank Anping Huang, Mingqiu Li, Jingdong Shao, Dianwei Wang, and Qi-Shu Yan for helpful discussions. This work is supported by the China Postdoctoral Science Foundation under Grant No. 2021M703169, the Fundamental Research Funds for the Central Universities E2E46303X2, the National Natural Science Foundation of China (NSFC) Grant Nos:12235016, 12221005, 11725523, 11735007, 12275108, and the Strategic Priority Research Program of Chinese Academy of Sciences under Grant Nos XDB34030000 and XDPB15, the start-up funding from University of Chinese Academy of Sciences(UCAS), the Fundamental Research Funds for the Central Universities, and the Guangdong Pearl River Talents Plan under Grant No. 2017GC010480.
\end{acknowledgments}

\bibliographystyle{apsrev}
\bibliography{ref}

\end{document}